\documentclass[fleqn,usenatbib]{mnras}

\usepackage{newtxtext,newtxmath}

\usepackage[T1]{fontenc}

\DeclareRobustCommand{\VAN}[3]{#2}
\let\VANthebibliography\thebibliography
\def\thebibliography{\DeclareRobustCommand{\VAN}[3]{##3}\VANthebibliography}

\usepackage{graphicx}	
\usepackage{amsmath}	

\graphicspath{{./figures/}} 
\usepackage{xcolor}
\usepackage{subfig}
\usepackage{hyperref}
\usepackage{ulem}

\title[Optimising emission line analysis in NGC~613]{Optimising the analysis of emission lines in galaxies: the case of the MUSE TIMER galaxy NGC~613}

\author[L. A. Silva-Lima et al.]{Luiz A. Silva-Lima,$^{1,2}$\thanks{E-mail: luiz.sl@outlook.com (LASL)}
Dimitri A. Gadotti,$^{2}$
Lucimara P. Martins,$^{1}$
Tutku Kolcu,$^{3,4}$
Paula R. T. Coelho,$^{5}$
\newauthor
Francesca Fragkoudi,$^{6}$
Taehyun Kim,$^{7}$
Camila de Sá-Freitas,$^{8}$
Jesús Falcón-Barroso,$^{9,10}$
\newauthor
Adriana de Lorenzo-Cáceres,$^{9,10}$
Jairo Méndez-Abreu,$^{9,10}$
Justus Neumann,$^{11}$
\newauthor
Miguel Querejeta,$^{12}$
and Patricia Sánchez-Blázquez,$^{13,14}$
\\
$^{1}$Núcleo de Astrofísica, Universidade Cidade de São Paulo, Rua Galvão Bueno, 868, São Paulo, Brazil\\
$^{2}$Centre for Extragalactic Astronomy, Department of Physics, Durham University, South Road, Durham DH1 3LE, UK\\
$^{3}$Astrophysics Research Institute, Liverpool John Moores University, IC2 Liverpool Science Park, 146 Brownlow Hill, L3 5RF, UK \\
$^{4}$School of Physics and Astronomy, University of Nottingham, University Park, Nottingham NG7 2RD, UK \\
$^{5}$Universidade de São Paulo, Instituto de Astronomia, Geofísica e Ciências Atmosféricas, Rua do Matão, 1226, 05508-090, São Paulo-SP, Brazil\\
$^{6}$Institute for Computational Cosmology, Department of Physics, Durham University, South Road, Durham DH1 3LE, UK\\
$^{7}$Department of Astronomy and Atmospheric Sciences, Kyungpook National University, Daegu 41566, Republic of Korea\\
$^{8}$European Southern Observatory, Alonso de Córdova 3107, Vitacura, Región Metropolitana, Chile\\
$^{9}$Departamento de Astrofísica. Universidad de La Laguna, Av. del Astrofísico Francisco Sánchez s/n, E-38206, La Laguna, Tenerife, Spain\\
$^{10}$Instituto de Astrofísica de Canarias. C/ Vía Láctea, S/N, E-38205 La Laguna, Tenerife, Spain\\
$^{11}$Max-Planck-Institut f\"{u}r Astronomie, K\"{o}nigstuhl 17, D-69117 Heidelberg, Germany\\
$^{12}$Observatorio Astronómico Nacional, C/ Alfonso XII 3, Madrid 28014, Spain\\
$^{13}$Departamento de Física de la Tierra y Astrofísica, Universidad Complutense de Madrid, E-28040 Madrid, Spain\\
$^{14}$Instituto de Física de Partículas y del Cosmos (IPARCOS), Universidad Complutense de Madrid, E-28040, Spain
}
\date{Accepted 2025 May 15. Received 2025 May 14; in original form 2025 January 20}

\pubyear{2025}

\begin{document}
\label{firstpage}
\pagerange{\pageref{firstpage}--\pageref{lastpage}}
\maketitle

\begin{abstract}
Galaxy evolution is driven by spatially distributed processes with varying timescales. 
Integral field spectroscopy provides spatially-resolved information about these processes. 
Nevertheless, disentangling these processes, which are related to both the underlying stellar populations and the interstellar medium can be challenging. 
We present a case study on NGC~613, observed with MUSE (Multi-Unit Spectroscopic Explorer) for the TIMER (Time Inference with MUSE in Extragalactic Rings) project, a local barred galaxy, which shows several gas ionisation mechanisms and is rich in both large and inner-scale stellar structures.
We develop a set of steps to overcome fundamental problems in the modelling of emission lines with multiple components, together with the characterisation of the stellar populations. 
That results in the disentanglement of the gas ionisation mechanisms and kinematics, along with an optimal parametrisation for star formation history recovery. 
Our analysis reveals evidence of gas inflows, which are associated with the bar dust lanes traced with \textit{Hubble} Space Telescope (HST). 
In addition, we show the gas kinematics in a central biconical outflow, which is aligned with a radio jet observed with Very Large Array (VLA). 
The emission line provides estimates of electron density, gas-phase metallicity, and the mass outflow rate, allowing us to distinguish intertwined ionisation mechanisms and to identify a part of the multiphase gas cycle in NGC 613. 
It traces the gas kinematics from the bar lanes to inner scale gas reservoirs, where it can eventually trigger star formation or AGN activity, as observed in the outflow.
\end{abstract}




\section{Introduction}

With the advent of integral field units (IFU), mounted on large telescopes, novel approaches for studying the star formation and the conditions of the interstellar medium (ISM) in galaxies were made possible.
The spatial resolution provided by the modern IFU is fundamental to trace the spatial variations of ISM properties like electron density, metallicity, kinematics, but also stellar population properties and structures.
Moreover, feedback from distinct gas excitation mechanisms like stellar phenomena, active galactic nuclei (AGN), and shocks can substantially impact the ISM and lead to spatial variations in its properties.
The characterisation of the emission lines produced by these sources is fundamental to understand the environment where they are generated, as well as the mechanisms responsible for the excitation of the gas.

Stellar structures also play an important role in the dynamics of gas in a galaxy, and acting in the secular evolution of other stellar structures \citep{sellwood1993, kormendy2004, sellwood2014}.
In particular, in late-type galaxies, a non-exhaustive list of stellar structures includes: discs (thick, thin, and nuclear), bulges, bars (with X-shaped and box/peanut), spiral arms, rings (nuclear, inner and outer), and lenses \citep[see][for a comprehensive morphological classification of nearby galaxies]{buta2015}.
In galaxies rich with these large and small, inner-scale substructures, the gravitational potential associated with these systems is often complex.
The interaction of gas with this potential may lead to particular implications in the gas dynamics.

Regarding the gas dynamics implications, one of the most remarkable stellar structures is the galactic bar (see e.g. \citealp{athanassoula1992} and \citealp{sormani2015} for the effects of stellar bars on gas dynamics).
Bars are common structures in spiral galaxies, being present in $\sim1/3$ of local galaxies, with the fraction increasing up to $\sim2/3$ when weak bars are included \citep[e.g.][]{eskridge2000, menendez-delmestre2007, buta2015}. 
The fraction of barred galaxies is also significant at higher redshifts, predicted in theoretical studies \citep{rosas-guevara2022, fragkoudi2024} and quantified by observations \citep{guo2023, guo2024, mendez-abreu2023, leconte2024}.
The non-axisymmetric potential of a bar exerts torques in the stellar components and the ISM, promoting momentum removal within the corotation radius, with bars holding ``galactic rivers'' \citep{sormani2023}, where the gas migrates towards the centre through the bar dust lanes.

This momentum exchange induced by bars led to a number of studies in the last few decades on the connection between bars and AGN activity, with divergent conclusions.
A body of evidence suggests that bars can favour the feeding of AGN or act in the building of a gas reservoir in the centre of galaxies \citep{knapen2000, laine2002, p.coelho2011, m.s.alonso2013, santiagoalonso2014, santiagoalonso2018, melaniegalloway2015, silva-lima2022, garland2024}, but other studies have found no support for a bar-AGN relation \citep{regan1999, martini2003, mauriciocisternas2013, goulding2017}.
This analysis is challenging due to the different timescales involved, the bar and AGN detection, and to the presence of dynamical resonances in the inner galactic scales.

At sub-kiloparsec scales, the gas can be trapped in resonances produced by the bar, preventing or delaying the AGN feeding process \citep{piner1995, regan1999, berentzen2007, shin2017}.
These regions are the locus of development of nuclear stellar structures like nuclear discs, where intense episodes of star formation take place.
These nuclear stellar structures can be studied in detail from IFU observations \citep[][]{falcon-barroso2006, delorenzo-caceres2013, delorenzo-caceres2019,  dimitria.gadotti2019, bittner2020, rosado-belza2020,pinna2021,martig2021}, with the relation between the large scale bar and inner structures development leading to methodologies that allow the bar age dating \citep{desa-freitas2023a,desa-freitas2023,sa-freitas2025}.

IFU observations also reveal the signs of feedback in the ISM, with several sources of ionisation at play concurrently \citep[e.g.][]{davies2017}.
AGN feedback, with spatially resolved ionisation cones was observed \citep{venturi2018, venturi2021, mingozzi2019, ruschel-dutra2021, bianchin2022}, thus allowing the characterisation of AGN activity, timescale and efficiency.
Stellar-driven processes are also prominent, with massive, young stars emitting ultraviolet radiation, which ionises the surrounding gas.
In addition to the ionisation by young stars, there is growing recognition of the role of evolved stellar populations in ionising the ISM.
Specifically, hot low-mass evolved stars can contribute to the ionisation of the gas \citep[][and references therein]{binette1994, macchetto1996, ho2008, cidfernandes2011, sarzi2010, stasinska2022}.

In the inner regions of galaxies there is fast evolution, with a combination of many of these different phenomena at play: star formation, shocks, gas accretion into the central region, inflows, and outflows of ionised gas.
The intertwined relation between these various components -- stellar structures and their populations, the gas-phase of the ISM, and the AGN activity --, produces many of the observable properties of galaxies.
However, disentangling all these effects spaxel-by-spaxel in a spatially resolved datacube requires a careful procedure to avoid both over-interpreting or misinterpreting the results, while extracting as much information as possible.
A detailed analysis of both the stellar continuum and the emission lines of a galaxy spectrum, in a consistent way, provides a powerful means for investigating the physical conditions within galaxies.

In this work, we present a detailed methodology for the interpretation of spectral datacubes of the central region of galaxies, where many dynamical and excitation mechanisms are at play. 
For the presentation of this methodology we use NGC~613, which provides an excellent case study since it is known to harbour both star-formation, an active nucleus, and it has a dynamically complex nuclear gaseous environment.

We combine a stellar continuum and an emission line model to reproduce spatially resolved observed spectra, and, at the same time, we characterise the star formation history and gas excitation mechanisms.
The star formation history is recovered through full-spectrum fitting techniques by combining single stellar population models.
The nebular emission lines are modelled with multi-component Gaussian profiles, with a model selection approach to determine the number of components required to minimise overfitting issues.
The stellar continuum and emission line measurements allow us to further derive properties like the gas and stellar kinematics, average stellar properties, dust reddening, ionisation mechanisms, gas-phase abundance, electron density, and to characterise aspects of the inflow and outflow of gas.
With this analysis, several potentially intertwined phenomena within the central kiloparsecs can be investigated, including, for example, the gas non-circular motions driven by stellar bars and the AGN feeding.
 
The paper is organised as follows: We introduce NGC~613 in Sect.~\ref{sect:ngc613}, briefly describe the data employed in Sect.~\ref{sect:data}, and in  Sect.~\ref{sect:stellar_popullation} we outline the stellar continuum fitting, followed by the emission line modelling in Sect.~\ref{sect:emission_fitting}. The star formation history and average stellar population properties are described in Sect.~\ref{sect:sfh}, and the gas-phase properties are described in Sect.~\ref{sect:gas_content}.
Finally, we discuss the results and their implications in Sect.~\ref{sect:discussion}, and summarise our study and main conclusions in Sect.~\ref{sect:conclusion}.
Throughout this paper we assume $\Lambda$CDM cosmology, with H$_0 = 70$~km~s$^{-1}$~Mpc$^{-1}$, $\Omega_m = 0.3$, $\Omega_{\Lambda}=0.7$ and $\Omega_r = 0.0$. 


\section{Introducing NGC~613}
\label{sect:ngc613}
\begin{figure}
\centering
\includegraphics[width=\columnwidth]{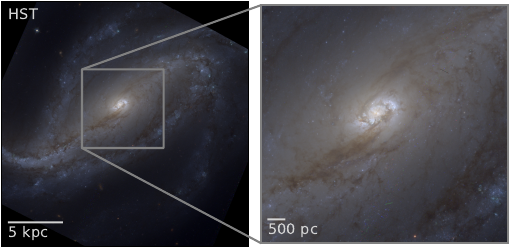}
\caption{NGC 613. Left: colour composite image from HST observations in the optical. Right: zoom-in region of the left panel, showing the observed field covered by the observations with MUSE. The image is composed with HST observations with the filters F435W, F606W, and F814W.}
\label{fig:ngc613}
\end{figure}

NGC~613 is a bright \cite[$M_B=-21.48\pm0.36~\rm{mag}$; \mbox{HyperLeda}\footnote{\url{http://leda.univ-lyon1.fr/}},][]{makarov2014}, local galaxy \citep[$d\approx26.4\pm5.3~\rm{Mpc}$;][]{nasonova2011}. 
It is also a massive galaxy, with $M_{\star}= 10^{11.1}~M_{\odot}$ \citep{sheth2010, munoz-mateos2013, querejeta2015} and a neutral hydrogen mass $M_{\rm{HI}}= 0.5\times10^{10}~M_{\odot}$ \citep[][estimated from the Hydrogen \mbox{21-cm} line flux measurements from \mbox{HyperLeda}]{dimitria.gadotti2019}.
The \textit{Hubble} Space Telescope (HST) colour composite image of NGC~613 is shown in Fig.~\ref{fig:ngc613}, with the filters F435W, F606W, and F814W from the archive observations (PID 15272; PI: Folatelli).

The galaxy is a barred spiral, classified as \mbox{SB($\rm{\underline{r}}$s,bl,nr)b} by \citet{buta2015}.
The inner ring is a large scale structure formed by wrapped spiral arms.
The barlens is thought to be a projection of the box/peanut, the latter being observed in edge-on disc galaxies, and understood as the inner part of the bar that has buckled and grown out of the disc plane \citep{laurikainen2014, athanassoula2015, athanassoula2016}.
In this classification, the innermost stellar structure is the nuclear ring and possibly associated with the bar inner Lindblad resonance (ILR) \citep[$\sim300~\rm{pc}$ in radius,][]{combes2019}, potentially interacting with the dust lanes in their contact points \citep{boker2008}.

Double-barred systems have been identified with IFU observations \citep[e.g.][]{delorenzo-caceres2019}. 
However, the existence of an inner bar in NGC~613 is controversial. 
The classification as a double-barred system was suggested by \citet{jungwiert1997}, based on ellipse fitting.
\citet{erwin2004} propose that this ``bar'' is actually an artifact produced by the star-forming ring.
Using $K$ band maps, \citet{boker2008} reinforces the hypothesis that the continuum is affected by star-forming clusters in the nuclear ring.

The overall star formation rate (SFR) of NGC~613 is estimated to be $5.3~M{_\odot}$~yr$^{-1}$ from infrared \citep{a.audibert2019}, and in the central region it is concentrated in the nuclear ring, where hot spots of young stellar populations have been observed \citep{boker2008}.
\citet{falcon-barroso2014} estimate the SFR ranges from 0.03 to $0.1~M_{\odot}$~yr$^{-1}$ in the ring, presenting evidence that supports the 
\textit{pearls on a string} scenario.
In this interpretation, gas flows through dust lanes connected to the bar, providing over-density regions in the ring which trigger star formation \citep{boker2008}.
Internal to the ring, SFR drops to $0.015~M_{\odot}$~yr$^{-1}$ in the nucleus, despite there resides a large amount of neutral cold gas of $\sim8
\times10^7~M_{\odot}$ \citep[]{falcon-barroso2014}.

Regarding AGN activity, the galaxy is classified as composite, having a Seyfert-like nucleus coexisting with starf-forming regions \citep{veron-cetty1986}.
Evidence for an AGN comes from the detection of the high ionisation line [\ion{Ne}{V}]$\lambda14.32~\rm{\mu m}$ \citep[97.1 eV;][]{goulding2009} and the H$_2$~(1–0)~S(1)/Br$\gamma$ ratio in the nuclear region in the mid-IR \citep{falcon-barroso2014}.
In the optical, the activity is pointed out as a combination of a low-ionisation nuclear emission line region (\mbox{LINER}) and star-forming \citep{dimitria.gadotti2019} according to the emission line ratios in the BPT diagnostic diagram \citep{baldwin1981}.
\citet{davies2017} identify regions dominated by distinct ionisation mechanisms employing the ratios [\ion{N}{II}]/H$\alpha$ and [\ion{S}{II}]/H$\alpha$, spatially resolved, as a proxy for a sequence of ionisation, spanning from star-forming regions to those with predominant AGN, shocks, or post-AGB.
On the other hand, from X-ray observations, the nucleus of NGC~613 is associated with a Compton-thick AGN \citep{castangia2013}.

NGC~613 also contains a $\rm{H_2O}$ megamaser \citep{henkel1984, braatz1996, zhang2006, farhan2023}.
It is suggested that this maser is associated with an observed radio jet \citep{kondratko2006, castangia2007}.
The position and luminosity \citep[$L_{\rm H_2O}\approx35~L_\odot$;][]{castangia2007} of the megamaser emission point to a potential association between the maser and the AGN activity of the galaxy \citep{falcon-barroso2014}.

The innermost structure identified is a two-arm trailing gas spiral ($r \lesssim150~\rm{pc}$) in the nuclear disc \citep{combes2019,a.audibert2019}.
This spiral was traced by \mbox{CO(3-2)} emission, observed with the Atacama Large Millimeter/submillimeter Array (ALMA), and was suggested to play a role in angular momentum transport.
With high spatial resolution provided by ALMA, \citet{a.audibert2019} observed filamentary structures linking the nuclear ring and the nuclear spiral.
The authors also modelled the torque due to the stellar potential, concluding it favours the gas to lose angular momentum and migrate from the ring to the centre, potentially feeding the AGN.

Extended nebular emissions are also observed in NGC~613, identified both with photometry and spectroscopy in the optical and also with potential counterpart observed from radio interferometry.
With narrow bands around H$\alpha$-[\ion{N}{II}] and [\ion{O}{III}], asymmetric elongations were revealed with position angle (PA) of $140\degr$ \citep{hummel1987}, where the nuclear disc is located.
At the same time, a structure alluding to an outflow and traced by the [\ion{O}{III}] emission is seen.
From [\ion{O}{III}], the outflow was revealed with PA of $25\degr$ and projected length of $\sim10\arcsec$ (1~kpc), and with evidence of a counterpart in almost the opposite direction \citep{hummel1987}.
A radio jet was observed with the Very Large Array (VLA) at 4.9~GHz near the centre, with an extension of $\sim5\arcsec$ (500~pc), a PA of $6\degr$, and almost aligned with the [\ion{O}{III}] emission.
Along with the jet, there is also evidence for the counter jet with $\rm{PA} = 227\degr$ \citep{hummel1987}.

\section{Data}
\label{sect:data}
In this paper, we use data from the Time Inference with MUSE in Extragalactic Rings\footnote{\url{https://www.muse-timer.org/}} \citep[TIMER;][]{dimitria.gadotti2019} project, corresponding to the galaxy NGC~613.
TIMER includes the observation of 21 nearby galaxies, obtained with the Multi-Unit Spectroscopic Explorer \citep[MUSE;][]{bacon2010} mounted at the Very Large Telescope (VLT).

MUSE is an integral-field spectrograph, with an almost square field-of-view (FoV), covering approximately $1\arcmin\times1\arcmin$, and with a spatial sampling of about $0\farcs2\times0\farcs2$, resulting in around $90\,000$ spectra per pointing.
The average full width at half maximum (FWHM) of the line-spread function (LSF) is about 2.6~\AA, and the spectra have a linear sampling of 1.25~\AA, with the nominal spectral coverage spanning from 4750 to 9350~\AA. 

The observation campaign (ESO programme 097.B-0640 PI:~Gadotti) took place during ESO period 97 from April to September 2016.
The FWHM of the seeing during the observations was between $0\farcs8$ and $0\farcs9$.
The data was reduced using the MUSE pipeline \citep[version 1.6][]{weilbacher2012} and for further details we refer the reader to \citet{dimitria.gadotti2019}.

\section{Stellar continuum fitting}
\label{sect:stellar_popullation}

To model the stellar population, we fit single stellar population (SSP) models to the observations, employing the full-spectral fitting technique.
We adopt full-spectral fitting applying the Penalized PiXel-Fitting (\texttt{pPXF}, version 9.4.1) code \citep{cappellari2004, cappellari2017, cappellari2023}.
With \texttt{pPXF}, we combine SSP models and emission line templates convolved with kernels that model the line-of-sight velocity distributions (LOSVDs).
The code is robust against undersampling, with the convolution executed in the Fourier space.
In addition to SSP models, templates of low-order Legendre polynomials and of emission lines are also included.
The final model is the optimal-weighted combination of those templates that minimise the residual sum in the $\chi^2$, reflecting the best agreement between model and observation \citep{cappellari2004}.

\subsection{SSP}

We use the MILES \citep{vazdekis2015} model library of SSPs as stellar templates, with BaSTI isochrones \citep{pietrinferni2004, pietrinferni2006}, Kroupa Revised \citep{kroupa2001} initial mass function (IMF), and spectral stellar libraries from \cite{sanchez-blazquez2006} and \citet{coelho2005}.
The models are initially linearly sampled in a spectral, covering the MUSE spectral range approximately between $4750-7400$~\AA, and with a spectral resolution of approximately 2.51~\text{\AA} \citep{falcon-barroso2011}. 
They cover a parameter space with varying age, metallicity, and [$\alpha$/Fe], producing a regular grid of models.
The age axis covers $0.03 - 0.10$~Gyr ($\Delta t=0.01$~Gyr), $0.10 - 0.50$~Gyr ($\Delta t=0.05$~Gyr), $0.50 - 1.00$~Gyr ($\Delta t=0.05$~Gyr), $1.00 - 4.00$~Gyr ($\Delta t=0.25$~Gyr), and $4.00 - 14.00$~Gyr ($\Delta t=0.50$~Gyr).
In line with the procedure adopted by \citet{desa-freitas2023}, we remove SSPs with \mbox{[M/H] < -1}, to include only stellar templates within the MILES safe range \citep{vazdekis2010,vazdekis2015}.
The remaining templates have the following metallicities: $-0.96, -0.66, -0.35, -0.25, 0.06, 0.15, 0.26$, and $0.40$.
With that, we also try to reduce the possibility of getting contributions from young and very metal-poor SSPs, a result that is commonly obtained in full-spectrum fitting, and might be the result of degeneracies or absence of templates with younger populations \citep{coelho2009, carrillo2020, bittner2020, e.emsellem2021}.
Finally, for the $[\alpha/\rm{Fe}]$ axis, we have two parameters for \mbox{$\alpha$-enhancement}: solar ($[\alpha/\rm{Fe}]=0.00$) and super-solar ($[\alpha/\rm{Fe}]=0.40$), which allow us to interpolate between the two most extreme values, providing an estimation of the $\alpha$-enhancement.

\subsection{Pre-processing of the observed spectra}
\label{sect:prepare_observations}

In this section, we describe the steps in preparing the observations for the spectral fitting process.
Before comparing the stellar models with the observations, some previous treatments are done, intended to remove observational effects.

The datacube is brought to the rest-frame using a redshift value of $z=0.004951$ \citep{meyer2004} and the spectra are corrected for the Galactic extinction.
For the correction, we use the dust maps provided by \citet{schlegel1998} and handled through the \texttt{dustmaps}\footnote{\url{https://dustmaps.readthedocs.io/en/latest/index.html}} \citep{green2018}.
Then, the observed spectral density fluxes are corrected for the wavelength dependent extinction with \citet{cardelli1989} law\footnote{Using the \texttt{extinction} package \url{https://extinction.readthedocs.io/en/latest/index.html}} and $R_V = 3.1$.

The information that can be extracted from the observations depends on a minimum signal-to-noise (S/N), measured per spectral element unless otherwise stated.
Recovering kinematics or star formation history may require different minimum S/N thresholds.
For stellar kinematics, the  radial velocity ($V_{\star}$) and velocity dispersion ($\sigma_{\star}$) were measured in the literature with S/N $\sim10$ \citep{westfall2019}, however, higher levels can be used in trying to improve accuracy, e.g. CALIFA \citep{sanchez2012} with S/N $\sim20$ and in \citet{bidaran2020} with $\sim40$.
For LOSVD modelled with a Gauss-Hermite quadrature, with higher-order moments that account for asymmetries in the LOSVD profile, larger S/N levels may be desirable.
Examples of determining kinematics with higher order moments $h_3$ and $h_{4}$ are: ATLAS$^{\rm{3D}}$\citep{cappellari2011}, Fornax3D \citep{sarzi2018}, and TIMER \citep{dimitria.gadotti2019}, all employing S/N $\sim40$.
Some studies include $h_{5}$ and $h_{6}$ as well, e.g. \citet[][S/N $\sim130$]{krajnovic2015} and \citet[][S/N $\sim70~\text{\AA}^{-1}$]{thater2022}.
On the other hand, for the recovery of the star formation history (SFH), TIMER applies S/N $\sim40$ \citep{dimitria.gadotti2019} and $\sim100$ \citep{bittner2021}, although $\sim50~\text{\AA}^{-1}$ is adopted by \citet{guerou2016}, PHANGS \citep{e.emsellem2021} employ S/N$\sim35$, and \citet{neumann2022} use S/N$\sim10$.

When the S/N per pixel is not sufficient, some kind of binning is necessary.
To achieve the desired level, we employ tessellation with Voronoi algorithm \texttt{VorBin} \citep{cappellari2003} with its default configuration, which includes Weighted Voronoi Tessellation \citep{diehl2006}.
We measure the signal and the uncertainty to estimate the S/N in the continuum following the approach in \citet{guerou2016}, using the spectral window $\lambda\lambda=5450-5550~\text{\AA}$. The reason for this choice comes from the fact that the stellar component in this region is less contaminated by sky and nebular emission.

The neighbour spectra were coadded in bins targeting S/N=100~pixel$^{-1}$, where we do not include corrections for spatial correlation.
In addition, the inclusion of noisy observations does not always significantly improve the S/N of a bin, but solely its size. 
For that reason, spaxels with S/N $<3$ were excluded.
In this way, we intend to have a uniform S/N throughout the FoV, thus being able to estimate the properties of the stellar populations which includes their kinematics up to the fourth-order Gauss-Hermite series, and the recovery of the star formation history. 

Finally, we normalise the flux in the observations.
We distinguish between two approaches to normalise the observations: scalar and field normalisation.
In the scalar mode, the normalisation factor is a single number corresponding to the global average of the observational flux, in a given spectral window, for all observed spectra.
The field normalisation approach provides a scalar field, where for each spectrum there is a normalisation factor resulting from the average of the flux individually measured.

The two normalisation techniques yield data with distinct characteristics, each of which carries significant implications. The scalar mode preserves the difference in flux between regions, such as between the centre and the outskirts of the FoV, while the field mode provides a more uniform flux level throughout the field.
Here, we employ the scalar field normalisation (as dissussed Sect.~\ref{subsec:stel_pop_param}).

\subsection{Pre-processing of the templates}
\label{sect:prepare_templates}

The SSPs spectral resolution is degraded to match the LSF of the observations.
We apply a convolution with the injection of the relative LSF.
The relative LSF is defined as the quadratic difference between MUSE LSF and the spectral resolution of the MILES templates (fixed at $\approx2.51~\text{\AA}$, meaning variable resolving power R).
The MUSE LSF as a function of wavelength is modelled with a polynomial regression by \citet[][Equation~8]{bacon2017}.
The templates have a smaller R than MUSE in a portion of the wavelength range, $\lambda\lambda\approx6670-7400~\text{\AA}$, and in such situation we truncate the relative LSF to zero in the concerned spectral region.

We also correct the MUSE LSF for redshift effects despite the object being a nearby galaxy \citep[][Sect.~2.4 for discussion\footnote{Code examples at \url{https://github.com/micappe/ppxf_examples}}]{cappellari2017}.
In this case, the effect is negligible, but might represent a significant bias for higher redshift galaxies.
The SSP templates were then log-rebinned to the same dispersion sampling of the observations. The individual fluxes of each SSP are conserved during the rebinning using \texttt{SpectCube}\footnote{\url{https://pypi.org/project/spectcube/}}.

To set the SSP weighting and avoid numerical issues, the SSPs are normalised, using a single scale.
The normalisation scale is the global flux average of the templates in the fitted range.
With this normalisation, the contribution of the SSPs in the fitting processes is light-weighted.

\begin{table}
\centering
\caption{Emission lines included in the fitting.}
\label{tab:emission_lines}
\begin{tabular}{cccc}
\hline
\multicolumn{1}{c}{Name} &
  \begin{tabular}[c]{@{}c@{}}Wavelength\\ {[}$\text{\AA}${]}\end{tabular} &
  \multicolumn{1}{c}{\begin{tabular}[c]{@{}c@{}}Ionisation\\ {potential [}eV{]} \end{tabular}} &
  \multicolumn{1}{c}{Constraints} \\ \hline
\multicolumn{4}{c}{Low-ionisation group}                        \\ \hline
H$\beta$         & 4861.33 & 13.60  & -                                 \\
H$\alpha$        & 6562.82 & 13.60  & -                                 \\
{[}\ion{N}{i}{]}        & 5197.90 & -      & -                                 \\
{[}\ion{N}{i}{]}        & 5200.26 & -      & -                                 \\
{[}\ion{O}{i}{]}        & 6300.30 & -      & -                                 \\
{[}\ion{O}{i}{]}        & 6363.78 & -      & 0.33{[}\ion{O}{i}{]}$\lambda$6300        \\
{[}\ion{N}{ii}{]}       & 6548.05 & 14.53  & 0.33{[}\ion{N}{ii}{]}$\lambda$6584       \\
{[}\ion{N}{ii}{]}       & 6583.46 & 14.53  & -                                 \\
{[}\ion{S}{ii}{]}       & 6716.44 & 10.36  & > 0.41{[}\ion{S}{II}{]}$\lambda$6731     \\ 
                 &         &        &  < 1.45{[}\ion{S}{ii}{]}$\lambda$6731   \\ 
{[}\ion{S}{ii}{]}       & 6730.82 & 10.36  & -                                 \\ \hline
\multicolumn{4}{c}{High-ionisation group}                         \\ \hline
{[}\ion{O}{iii}{]}      & 4958.91 & 35.12  & 0.33{[}\ion{O}{iii}{]}$\lambda$5007      \\
{[}\ion{O}{III}{]}      & 5006.84 & 35.12  & -                                 \\
\ion{He}{i}             & 5875.62 & 25.58  & -                                 \\ \hline
\end{tabular}
\end{table}

In the spectral window we used for the fit, H$\beta$ is the absorption-line with the highest sensitivity to the stellar population age.
Due to a combination of ionisation sources, the nebular emission of the recombination lines is ubiquitous in the FoV.
Thus, most of the spectra present a superposition of emission and absorption line in the H$\beta$ region.
Masking this line could result in a reduction in the ability to distinguish between SSPs of different ages. We thus decided to model the emission lines along with the stellar population.

In Table~\ref{tab:emission_lines}, we present the set of emission lines included in our model, used for the stellar population and later for the emission line fitting with multiple components  (Sect.~\ref{sect:emission_fitting}).
The emission line templates have the initial width that matches the MUSE LSF, thus the measured velocity dispersions correspond to the intrinsic value. 
Adopting the same procedure used in PHANGS \citep{e.emsellem2021}, we divide lines into groups according to the ionisation potential.
For each group, a single LOSVD is employed.

Gauss-Hermite quadratures are used to parametrise the LOSVD \citep{vandermarel1993, gerhard1993} both for emission lines and stellar population.
This parametrisation is intended to take into account asymmetries that result in departures from a Gaussian LOSVD.
The fourth-order quadrature have moments ($V_{\star}, \sigma_{\star}, h_3, h_4$) with different physical meaning.
The first two correspond to the radial velocity and velocity dispersion, while $h_3$ and $h_4$ are related, respectively, to the skewness and kurtosis \citep{gadotti2005}.

\subsection{Fitting procedure}
In this section, we describe the steps used in the data fitting pipeline.
Here the stellar continuum is modelled along with the emission lines, with the latter being modelled using Gauss-Hermite polynomials. 
This first approach captures the complexity of the emission line profiles allowing a robust analysis of the stellar continuum.
In the next section we reanalyse the emission lines in more depth, employing multiple Gaussian components.
The procedures adopted here were inspired by previous stellar population fitting procedures described in the literature.
It was majorly influenced by the MaNGA DAP \citep{belfiore2019, westfall2019}, TIMER survey \citep{dimitria.gadotti2019} with the GIST pipeline \citep{bittner2019}, and PHANGS-MUSE survey \citep{e.emsellem2021}.
To work with the large dataset, we take advantage of the support for memory mapping provided by NumPy \citep{harris2020}. In addition, we employ the Message Passing Interface (MPI) tools provided by \texttt{mpi4py} \citep{dalcin2005, dalcin2008, dalcin2011, dalcin2021} to distribute the process across computer nodes.

The fitting process encompasses four steps: mask optimisation, extinction fitting, kinematics fitting and stellar population history recovery.
Every bin yielded during the tessellation is processed independently in the sequence described in the following sections.

\subsubsection{Optimise spectral masks}

On this first stage, we optimise the spectral mask. 
Initially, only the start and end of the spectral axis were masked.
Following \citet{westfall2019}, we mask these regions to avoid artefacts during the fit procedure that can be produced due to the convolution with a kernel.
This kernel is going to have the width equal to the assumed maximum stellar velocity dispersion.
Considering the velocity dispersion to be $\sigma_{\star}\approx200~\rm{km~s}^{-1}$, we employ masks with width $\pm3\sigma_{\star}$.

Next, we produce a new mask, for each spectrum, that will then be used throughout the fitting.
First, we conduct a fit of the SSPs, emission lines, and corresponding LOSVDs. 
In addition, an 8th order additive Legendre polynomial is included to take into account differences of flux calibration between observation and SSPs.
Using the residuals from this fit, we employ a sigma-clipping algorithm provided by \citet{cappellari2023} for masking bad pixels, cosmic rays or artefacts that might remain after the reduction. 
With this optimisation, we do not remove the initial mask, but we include new points of the spectral axis to be masked by combining the initial mask with the newly generated one in the iterative process with sigma-clipping.

\subsubsection{Internal extinction fitting}
\label{sect:stellar_dust_fitting}

We also employ \texttt{pPXF} for internal dust extinction fitting.
The spectra were corrected for foreground Galactic extinction Sect.~\ref{sect:prepare_observations}, but there is also an additional extinction effect from the dust within the galaxy. 
Here, we aim to disentangle this effect from the stellar population fitting. 
The procedure is analogous to the one adopted in the PHANGS-MUSE survey \citep{e.emsellem2021}.
At this stage, we do not employ Legendre polynomials, to avoid any potential degeneracy between the polynomial and the extinction curve.
The extinction curve adopted is parameterised by the Calzetti's law \citep{calzetti2000} with $R_V = 4.05$.
The LOSVDs and template weights are free to vary. 
As a result, we obtain the best fit value of $A_V$ for the stellar extinction.
However, later, on Sect.~\ref{sect:gas_content}, we discuss the measurements for nebular extinction based on the Balmer decrement.
After deriving $A_V$, we proceed with the correction in a step prior to proceeding with the subsequent fitting stage, using the \texttt{extinction} package and propagating that to the uncertainty spectrum.

\subsubsection{Stellar kinematics}

The subsequent step is the kinematics fitting, inferring the LOSVDs moments using the whole spectral axis interval of the observations which is covered by the SSP models, namely, $\approx4750-7400$~\AA.
Here, we employ an 8th order additive Legendre polynomial to avoid template mismatch and take into account differences in calibration.
The choice for the additive polynomial results in a faster execution than the multiplicative one.
Along with the polynomials and templates, we employ the observed spectrum corrected for reddening in the previous step, with both gas and stellar LOSVDs being parametrised by a four-moment Gauss-Hermite quadratures, where the higher order moments $h_3$ and $h_4$ reflect deviations from a pure Gaussian LOSVD, which can be caused, e.g., due to the superposition of stellar populations with distinct kinematical properties or triaxial structures.

\subsubsection{Stellar population fitting with regularised solution}
\label{subsec:stel_pop_param}

Finally, we fit the stellar population parameters, modelling the SFH.
We take advantage of the products of the previous steps, using the optimised mask, the extinction-corrected spectra, and also the fitted kinematics.
We now keep the stellar kinematics fixed, broadening the templates to avoid the metallicity-kinematics degeneracy \citep{koleva2008, sanchez-blazquez2011}.
Here, we include an eighth order multiplicative Legendre polynomial, in contrast with the additive one adopted previously, as the multiplicative polynomials result in a slower execution but prevent changes in the strength of spectral stellar features \citep{cappellari2017}.
This model is then fitted to the observed spectrum employing regularisation.
The SFH was fitted in a shorter interval, between 4800 and 5800~\AA, following \citet{bittner2021}.
This smaller range for stellar population fitting is chosen to avoid noise in the near IR due to telluric absorption corrections.
Also because strong emission lines are in this range, and, in some cases, their complex profiles are difficult to model in parallel to the continuum fit.
Finally, the recovery of the stellar population parameter can benefit from the short interval \citep{goncalves2020}.

\begin{figure*}
\includegraphics[width=\textwidth]{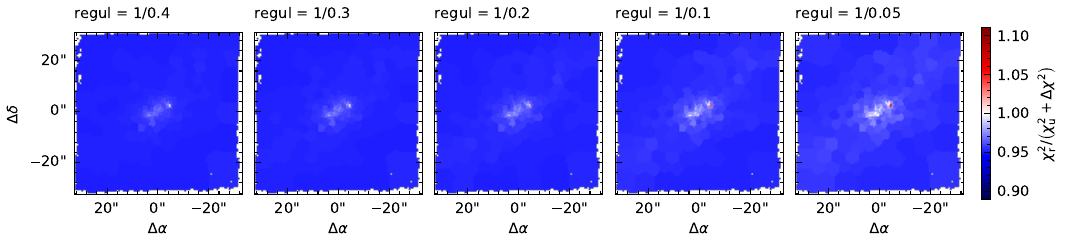}
\includegraphics[width=\textwidth]{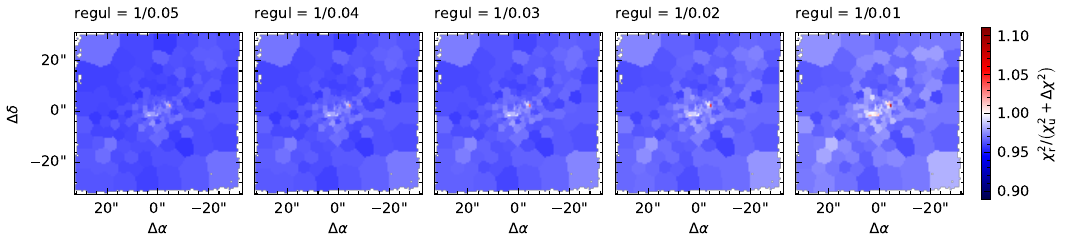}
\caption{Setting the regularisation parameter and the impact of the observation normalisation in the parametrisation. The regularisation is parametrised with regul=1/$\Delta$, where $\Delta$ is the regularisation parameter. The maps are colour coded according to the ratio between the $\chi_{{\rm reg}}$ of the fit and the maximum degraded $\chi_{{\rm unr}} + \Delta\chi$ of a regularised solution, regarding the criterion outlined in \citet{press2007}. Top row: observations normalised by a scalar factor, for $\Delta$ between 0.4 and 0.05. Bottom row: normalisation with scalar field (a scalar factor for each spectrum), with $\Delta$ varying from 0.05 to 0.01.}
\label{fig:regul}
\end{figure*}

This last stage concerns the star formation history recovery.
However, this is an inverse problem within the class of the ill-posed problems, with some assumption about the solution being required to deal with the degeneracies between age and metallicity.
At the same time, and due in part to the noise in the observations, there are many combinations of templates that may result in comparable values of goodness of fitting statistics.
These multiple noisy solutions, similar in terms of $\chi^2$, for example, may not be physically meaningful as an SFH, although some average properties, such as mean age and mean metallicity, can be the same. 
Different approaches have been used in literature for the star-formation history recovery, e.g. \texttt{STARLIGHT} \citep{cidfernandes2005}, \texttt{STECKMAP} \citep{ocvirk2006}, \texttt{VESPA} \citep{tojeiro2007}, and, more recently, \texttt{FIREFLY} \citep{wilkinson2017}.
Using \texttt{pPXF}, it is possible to estimate the SFH with a shrinkage method applied to the SSP weights, although, given its flexibility, different approaches can be employed in the code, as for example with resampling techniques \cite[see][for an application of Monte Carlo methods]{e.emsellem2021}.

In particular, the method we applied in \texttt{pPXF} is the Tikhonov regularisation (\citealp{tikhonov1977}; see also \citealp{cappellari2017} and references therein), parametrised in the code by the keyword $\texttt{regul} = 1/\Delta$, where the $\Delta$ parameter limits the second numerical derivative between the weights of neighbour templates in the SSP grid.
Thus, anything that affects the weights may impact the regularisation parametrisation.
The regularisation is set in such a way that we achieve a good balance in the bias-variance tradeoff.
With the increment of the regularisation the SFH assumes a smoother shape, and the fit becomes more robust against noise, reducing the variance in the solution due to perturbations of background signals.
The side effect is a more biased solution, with more subtle SFH features prone to be hidden during the fitting.
The other extreme of the bias-variance tradeoff is the unregularised fit, resulting in a noisy and occasionally unphysical SFH, susceptible to the observation noise.

Setting the regularisation parameter is then conditioned to data and model characteristics.
For example, the final values of the SSP weights are subject to the employed normalisation of the observed spectra as follows.
The continuum level in brighter central spaxels is generally considerably higher than the ones in the FoV outskirts, leading to large differences in the weights of neighbour SSPs in the grid of SSPs used. As a result, for the same regularisation parameter, a brighter spectrum would be over-regularised while an under-regularisation could be observed in fainter spectra.
In this case, the choice of a single regularisation parameter for IFU data would not be a trivial task, and only an intermediate value of $\Delta$ reasonably suitable for the whole FoV is possible.

In addition, one would expect the parameter to be conditioned by the number of templates in the SSP grid.
With a finer grid, the weights may be influenced since the contribution for models of a given interval of age and metallicity may be distributed across more templates, leading to smaller weights in comparison to a coarser grid.
So, when employing different SSP libraries, the regularisation of the fit may also be affected.
Nevertheless, the comparison of regularisation parameters for different libraries and the significance of the differences is out of the scope of this work.

To set a good balance in the parametrisation, we adopt the widespread recommendation to control the regularisation, limiting the degradation of the $\chi^2$ statistics \citep{press2007}.
The $\Delta$ parameter is reduced until we accomplish a solution with $\chi^2_{\rm{reg}} \approx \chi^2_{\rm{unr}} + \Delta\chi^2$, where $\chi^2_{\rm{reg}}$ is the statistics of the regularised solution, $\chi^2_{\rm{unr}}$ comes from the unregularised fit, and $\Delta \chi^2$ is given by the number of spectral pixels included in the fit, $\Delta\chi^2 = \sqrt{2N_{\rm{pixels}}}$ \citep{press2007, onodera2012}.

In Fig.~\ref{fig:regul}, we present a test for setting the regularisation of a fit using a given model and observation.
We compare the ratio of the obtained statistics ($\chi^2_{\rm{reg}}$) with the expected one ($\chi^2_{\rm{unr}} + \Delta\chi^2$).
To make the test more feasible, we sample the cube, keeping one element out of three in the right ascension axis and applying the same criteria to the declination axis, resulting in a representative cube of the original observation, with about one tenth of the spectra.
For each regularisation value, we conduct the fitting following the previous described steps.
In the upper row, we highlight the effect of the regularisation with the scalar normalisation of the observations (see Sect.~\ref{sect:prepare_observations}),
while in the bottom row, we find a more uniform regularisation given a $\Delta$ parameter, with the observations been normalised in the field mode (Sect.~\ref{sect:prepare_observations}).
This shows how, by using the field mode to normalise the observed spectra, we can employ regularisation obtaining a more optimal balance in the bias-variance trade-off much more uniformly and closer to unity across the whole FoV.
Using the field mode normalisation, we adopt a conservative value of $\Delta=0.05
$ to recover the SFH.
Nevertheless, even more constrained regularisation could be applied according to the tests.

\section{Emission line fitting with multiple components}
\label{sect:emission_fitting}

Multiple sources of gas excitation can be present in a galaxy, with a variety of spatial distributions.
The physical conditions of these regions are reflected in emission lines during its de-excitation.
Among such excitation mechanisms are those of stellar origin and those related to AGN.
Examples of stellar excitation mechanisms are young, massive stars, present in \ion{H}{II} regions and post-AGB stars, which can mimic a low-luminosity AGN or a LINER \citep[][and references therein]{ho2008, stasinska2022}.
On the other hand, an AGN with a hot accretion disc can have an even harder radiation field, producing stronger high-ionisation lines. 
These distinct excitation mechanisms also lead to characteristic emission line ratios and widths, allowing for the classification of the ionisation source and characterisation of the physical conditions of the gas.

The presence of kinematically distinct components in the stellar and nebular spectra can be observed through asymmetries in the absorption- and emission line profiles.
In the stellar LOSVDs, deviations from Gaussian profiles can be well-modelled with the inclusion of higher order moments in the Gauss-Hermite quadrature.
However, for the nebular emission, the Gauss-Hermite quadrature may not be sufficient to replicate the outcome of the emission line superposition of gas de-excitation originating from distinct regions, in particular in cases where the emission line sources have very different kinematics.
It is through the implementation of multiple Gaussian components, which are meant to model the kinematics of each component in the gas emission, that it is feasible to reproduce the spectra observed in galaxies with multiple ionisation sources.

One of the central problems in fitting emission lines with multiple components is related to the presence of local minima populating the parameter space, as shown by \citet{ho2016}.
As a consequence, the solution may depend on the initial guesses of the fitting, which is a critical step of the fitting process.
There are approaches for fitting multiple Gaussians available which avoid local minima.
In LaZy-IFU \citep[\texttt{lzifu};][]{ho2016}, for example, it is possible to fit both the continuum with \texttt{pPXF} and the emission lines with the Levenberg-Marquardt least-square implementation on \texttt{MPFIT} \citep{markwardt2009}.
The fit in \texttt{lzifu} is done by providing a combination of possible initial guesses, with the resulting $\chi^2$ of the fit being compared in order to avoid local minima.
The search for the global minima can also be conducted on the more recent version of \texttt{pPXF} \citep{cappellari2023} using the differential evolution\footnote{\url{https://docs.scipy.org/doc/scipy/reference/generated/scipy.optimize.differential_evolution.html}} algorithm implemented on SciPy \citep{2020SciPy-NMeth}, which is robust at the expense of being more computationally expensive.
Taking advantage of the physical spatial correlation between neighbour areas in a galaxy has resulted in the approach of \mbox{\texttt{IFSCUBE}} \citep{ruschel-dutra2021}, where the fitting process starts at a given position of the FoV (e.g. continuum peak) proceeding in a spiral pattern around it, using the previous fitting to obtain well-informed initial guesses, producing smoothed maps. Multiphase gas kinematics modelling is also conducted in ALMA submilimetre observations \citep{henshaw2019, schinnerer2023} using \texttt{scousepy} \citep{henshaw2019,henshaw2020}.

Other examples of using multiple Gaussian components include observations from \mbox{MAGNUM} \citep{venturi2018, venturi2021, mingozzi2019}, \mbox{SAMI} \citep{ho2014, medling2018, croom2021}, and also \citet{davies2016, davies2017}, \citet{kakkad2018}, \citet{smethurst2021}, and \citet{kolcu2023}.
The fit configurations are not homogenous in the literature, changing according to the scientific investigations, having distinct amplitude and kinematics constraints or fixed ratios between the emission lines.

\subsection{Fitting procedure}

We fit multiple Gaussian to the emission lines employing \texttt{pPXF}, combining this tool with some statistical methods in a few steps to obtain a spaxel-by-spaxel modelling, while also aiming for robustness against local minima.
The continuum is also modelled in order to reduce the effects of the stellar features in the emission line modelling (e.g. Balmer lines absorption).
Finally, we apply a model selection to distinguish between spatial regions
where a different number of components is required.

\subsubsection{Stellar component}

Before the emission line fitting, the stellar continuum is also taken into account. 
For this, we use the results from Sect.~\ref{sect:stellar_popullation} 

The underlying continuum component is modelled similarly to that described in Sect.~\ref{sect:stellar_popullation}, although here we only use the unregularised fits.
We also apply tessellation in this first stage, measuring the S/N in the vicinity of H$\beta$, in line with the approach adopted in \citet{venturi2018, mingozzi2019}, and targeting a S/N $\sim100$, since H$\beta$ is more attenuated due to dust in comparison to the other strong emission lines.

\subsubsection{Global optimisation}

The modelled emission lines are divided into two groups, as described in Table~\ref{tab:emission_lines}. 
These groups present distinct ionisation potential, and can trace different physical phenomena.

To search for the global minimum in the parameter space, we employ the differential evolution algorithm on \texttt{pPXF}.
This is an evolutionary algorithm: it creates a set of candidate solutions and iteratively refines them, producing new candidates until the conditions to stop are satisfied.
This algorithm is computationally costly, and so the applied tessellation has also the advantage of providing a reduced set of higher S/N spectra to execute the fitting.

Once we have the template for the underlying continuum, we proceed to model the emission, producing solutions with one, two, and three Gaussians for each emission line in each spaxel.
Initially, the components are organised according to the velocity dispersion, the narrowest line being the first component, in the same fashion as \citet{ho2016}.
Not necessarily this order will make sense in every case. In fact, we reorganise the order of the components, noting that in the region with 3 components, the additional component seems to have an intermediate velocity dispersion between the narrower and broader components.
Initial guesses are not required for this minimisation approach, and we only establish bounds for the radial velocities and velocity dispersions of the components, which reduces required assumptions about the solution.

\subsubsection{Spatially resolved emission line fitting}

\begin{figure}
\centering
\includegraphics[width=\columnwidth]{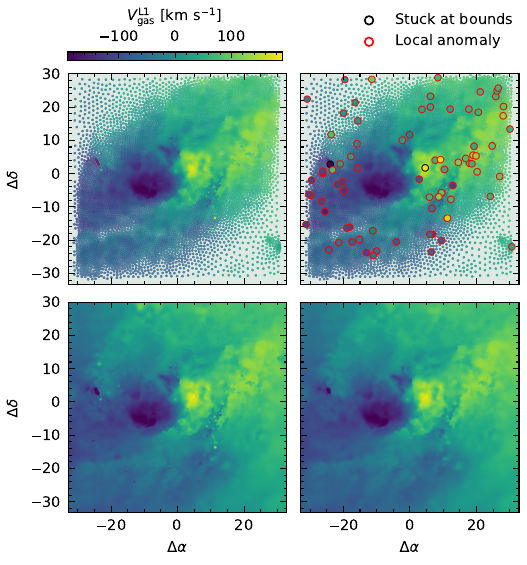}
\caption{Obtaining a solution from interpolation and outlier detection. In the upper panels, the radial velocity for a field with tessellation is displayed. The lower panels display the interpolation products by radial basis function. On the right, the data goes through an outlier detection process. The global optimisation may not achieve the minimum in some sparse bins and get stalled in local minima; we try to detect these bins comparing them with the neighbour bins (dots circle in red; upper right) so they are not propagated in the interpolation (lower panels).}
\label{fig:rbf}
\end{figure}

We use the previous binned solution to generate initial estimates for conducting a spaxel-by-spaxel, non-binned fit, where the observations were convolved with a minimal Gaussian kernel of $\sigma=1$~pixel ($0\farcs2$) in line with \citet{venturi2018}, and without significant loss of spatial resolution, but with an S/N increase as a result.
By commencing near the global minima, we expect to minimise the chances of becoming trapped in a local minimum.
To create the spatially resolved initial guesses, we use the radial basis function\footnote{\url{https://docs.scipy.org/doc/scipy/reference/generated/scipy.interpolate.RBFInterpolator.html}} (RBF) method with the function \textit{thin plate spline} as the base. 

Differential evolution provides robustness to find the global solution. 
Nevertheless, there is no guarantee that the global minimum will be found.
As an example of this process, we show on the upper left panel of Fig.~\ref{fig:rbf} the radial velocity of the low-ionisation group: some bins display very different values from their neighbours, and are potentially bins were the global optimisation failed to find the global minima, namely outliers/anomalous bins.
If we include these bins in the RBF interpolation to create initial guesses for the spaxel-by-spaxel solution, the outliers will be propagate and appear as uniformities in the FoV (lower left panel).
In order to minimise this effect, we employ an outlier/anomaly detection method.

Initially, we find the nearest neighbours of each bin using k-nearest neighbours and, next, employing the local outlier factor (LOF), both implemented on \texttt{scikit-learn}\footnote{\url{https://scikit-learn.org}} \citep{scikit-learn}. 
We train the model with these neighbours and compare with the reference object to detect potential outliers.
Fits which are stuck at bounds are also identified as solutions that have potentially failed.
The detection result is presented in the upper right panel of Fig.~\ref{fig:rbf}, where the outliers are circled in red.
After masking the outliers, we reapply the RBF interpolation, providing the initial guesses displayed in the lower right panel.

We apply the same interpolation method to create all initial parameters (line-of-sight velocity and velocity dispersion for each component), and next we fit the emission line with the minimisation done with the \texttt{CapFit} method \citep{cappellari2023}.
Given this solution is obtained spaxel-by-spaxel, differences between the continuum of the spectra and the continuum derived with the binned FoV may arise, so we include a second order multiplicative Legendre polynomial to fine-tune the continuum.
Once again, the components are organised according to the velocity dispersion, and we produce solutions up to three components for the whole FoV.

\subsection{Model selection}

We have fitted the spectra using different numbers of components for the emission lines (from 1 to 3) and, subsequently, it was possible to apply a selection criterion to identify the minimum number of components that best fits the data.
In this way, we produce three solutions, and then apply a model selection method to identify the number of components that result in the model which is simpler, and, at the same time, more explanatory.

We chose the Akaike information criterion \citep[AIC;][]{akaike1974} for model selection.
This criterion, rooted in information theory, enables the comparison of models to determine which one is more informative.
Thus, when comparing two models, one of which possesses greater complexity and more free parameters, we can assess whether the more complex model provides any additional level of information in its application.
The criterion is presented in \citet{burnham2004} as:
\begin{align}
    \rm{AIC} = -2\log(\mathcal{L}(\hat{\theta})) + 2k,
\end{align}
where $\mathcal{L}(\hat{\theta})$ is the model likelihood, with the estimated parameters $\hat{\theta}$, and $k$ is the number of free parameters in the model.
Alternatively, one can employ the formulation for a limited number of data points $n$ \citep{hurvich1989}:
\begin{align}
    \rm{AICc} = -2\log(\mathcal{L}(\hat{\theta})) + 2k + \dfrac{2k(k+1)}{n-k-1},
\end{align}
and thus:
\begin{align}
    \rm{AICc} = \rm{AIC} + \dfrac{2k(k+1)}{n-k-1}.
    \label{eq:aicc}
\end{align}
For large $n$ both values converge, thus we employ AICc to avoid any concern about the minimum number of data points.

We adopt the following procedure to compare the models and decide which one to employ. 
We use the AICc differences ($\Delta_i$), defined as \citep{burnham2004}:
\begin{align}
     \Delta_i =\rm{AICc}_i - \rm{AICc}_{min},
\end{align}
where $\rm{AICc}_{min}$ is the smallest $\rm{AICc}$. 
Since AICc values are relative and not absolute measurements, only the difference $\Delta_i$ is fundamental for the selection.
Then we compute the Akaike weights ($w_i)$, in the set $R$ of models:
\begin{align}
     w_i = \dfrac{\exp(-\frac{1}{2}\Delta_i)}{\sum_{r = 1}^{R} \exp(-\frac{1}{2}\Delta_r)}.
\end{align}

To indicate how significant the choice of one model is in relation to another, we use the evidence ratio $w_i/w_j$ of the $i$-th model in comparison to $j$-th one.
The evidence ratio provides a means of estimating the level of statistical support for choosing one model over another. An evidence ratio greater than 2.7 provides reasonable support for choosing a more complex model over a simpler one \citep[see][]{burnham2004}.
In addition, we require all the stronger lines in the low ionisation group, namely [\ion{N}{II}], [\ion{S}{II}], H$\alpha$ and H$\beta$, to have at least a detection level of 1$\sigma$ each in relation to the residual noise.
For the high ionisation group, we impose a higher level of $3\sigma$ on [\ion{O}{III}], since there are fewer lines in this group.

\begin{figure*}
\centering
\includegraphics[width=\textwidth]{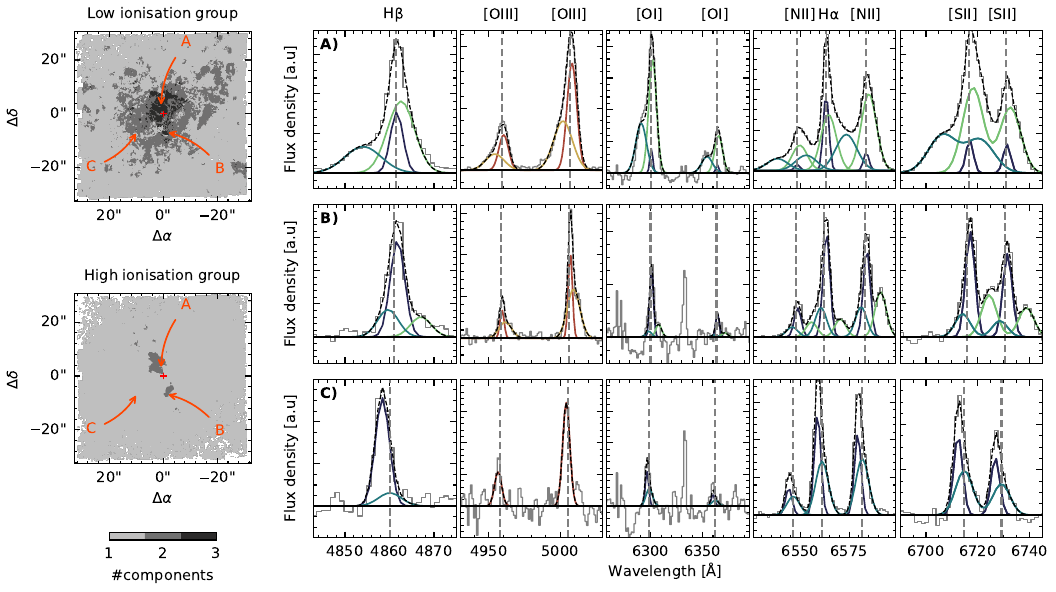}
\caption{Application of AIC to observations of NGC 613. Upper left: Application to low-ionisation lines. Lower left: Application to lines with high-ionisation. The maps are shown in shades of grey according to the number of components. The low ionisation group requires one component in the FoV outskirts and has up to three components in the FoV centre, while the high ionisation group exbibits up to two components, forming a bipolar pattern. In the left-hand panels, we highlight three regions labelled as A, B, and C, and we show the corresponding best fit emission line model for each one of these regions. In these panels, we zoom-in the spectra around strong emission lines. The Gaussian lines in the group of low ionisation are shown in teal shades, while the Gaussian in the group of high ionisation are shown in orange. The background grey histograms are the continuum-subtracted observations, while the dashed black curves are the best fits that result from the sum of the Gaussian components. Vertical dashed lines indicates the rest frame emission line wavelength, Doppler shifted according to the $V_{\star}$ measured in the corresponding spaxel and the horizontal continuous black lines indicate the zero-point of the relative flux.}
\label{fig:aic_selection}
\end{figure*}

We apply AICc to the two ionisation groups, with Equation~\ref{eq:aicc}.
As expected, it leads to different selections for the different groups.
As shown in Fig.~\ref{fig:aic_selection}, both groups have a higher number of components detected in the FoV centre, where S/N is higher and more deviations from a Gaussian profile can be perceived in the emission lines.
In addition, in the galaxy centre, we expect more overlapping stellar structures and ionisation soucers with distinct kinematics, and, consequently, emission line profiles that deviate from a pure Gaussian.
Nevertheless, with the high-ionisation group we can trace up to two components, while for the lower-ionisation up to three components could be noticed.
A possible explanation for this is the larger number of lines in the latter group, so that the detection of subtle secondary components  can be done more easily than in comparison with the high ionisation group.
Also, the high-ionisation line [\ion{O}{III}], which is the strongest line in its group, traces a distinct physical phenomenon, alluding to a biconical/bipolar pattern of an outflow.

With these detection levels adopted, we aim to have robust measurements, while being able to model secondary components that are usually less intense.
The selection criteria applied may result in reliable estimates of the LOSVD, however, additional procedures as done in \citet{sarzi2006}, associated with data characteristics, are necessary to assess model limitations and systematic errors.
For measurements relying on fluxes or line ratios that can be more sensitive to noise, e.g., the Balmer decrement, diagnostic diagrams, and metallicity, we impose an additional detection level of $3\sigma$ on the first components of all the lines, which mainly resulted in the masking of regions in the outskirts of the FoV.

\section{Stellar population properties}
\label{sect:sfh}

\subsection{Average stellar population properties}

\begin{figure}
\includegraphics[width=\columnwidth]{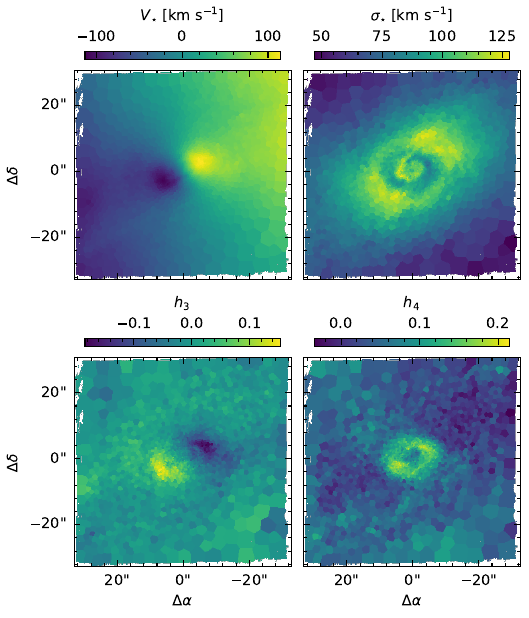}
\caption{Stellar kinematics, modelled as moments of Gauss-Hermite series. Top left: line-of-sight stellar velocity (${V_\star}$). Top right: stellar velocity dispersion ($\sigma_{\star}$). Bottom left: skewness ($h_3$). Bottom right: kurtosis ($h_4$).}
\label{fig:stellar_kin}
\end{figure}

From the fitting with \texttt{pPXF}, the average kinematics of stellar populations at the bin level can be modelled.
In Fig.~\ref{fig:stellar_kin} we show the four moments of the Gauss-Hermite series.
The $V_{\star}$ is typical of discs seen in projection.
On the other hand, the $\sigma_{\star}$ indicates a lower velocity dispersion in the nuclear disc identified by \citet{dimitria.gadotti2020}. 
The presence of this structure result in the anti-correlation between $V_{\star}-h_3$, which indicates a rotating disc \citep{dimitria.gadotti2019}, whereas the positive $h_4$ suggests the superposition of two stellar populations with distinct kinematics.
Furthermore, two regions with enhanced velocity dispersion are shown along the diagonal that coincides with the bar, and may indicate the existence of a box/peanut \citep[see the case of NGC~4643,][]{dimitria.gadotti2019}.

\begin{figure*}
\includegraphics[width=\textwidth]{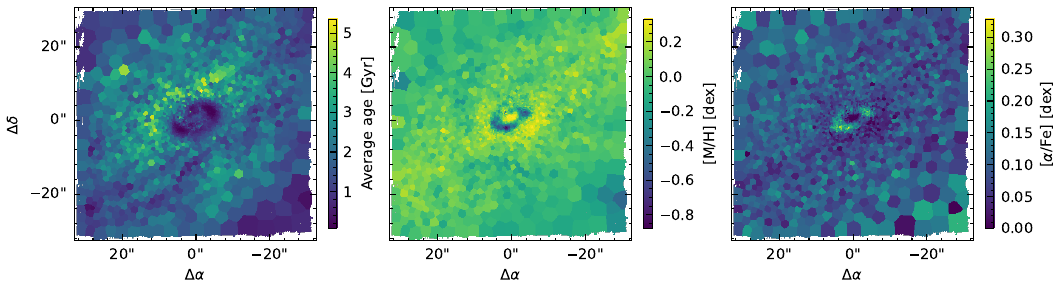}
\caption{Average light-weighted stellar properties. The panels show, respectively, the average age (left), [M/H] (middle), and [$\alpha$/Fe] (right).}
\label{fig:stellar_average_light}
\end{figure*}

In addition, as a result of the regularised fitting, we can estimate the average stellar properties spatially resolved.
In Fig.~\ref{fig:stellar_average_light} the light-weighted average age, metallicity and \mbox{$\alpha$-enhancement} of the stellar population are shown, according to Equations~\ref{eq:average_age}, \ref{eq:average_mh} and \ref{eq:average_afe}, respectively:
\begin{gather}
\label{eq:average_age}
    \langle \log_{10} t \rangle = \dfrac{\sum_i w_i ~ \log_{10} t_{\text{SSP}, i}}{\sum_i w_i} \\
\label{eq:average_mh}
    \langle \text{[M/H]} \rangle = \dfrac{\sum_i w_i ~ \text{[M/H]}_{\text{SSP}, i}}{\sum_i w_i} \\
\label{eq:average_afe}
    \langle \text{[$\alpha$/Fe]} \rangle = \dfrac{\sum_i w_i ~ \text{[$\alpha$/Fe]}_{\text{SSP}, i}}{\sum_i w_i} 
\end{gather}
where $w_i$ is the contribution from the \mbox{$i$-th} template with age $t_{\text{SSP}, i}$, metallicity $\text{[M/H]}_{\text{SSP}, i}$ and $\alpha$-enhancement $\text{[$\alpha$/Fe]}_{\text{SSP}, i}$. 

The age sampling of the SSP grid is neither linear nor logarithmic. Our choice of computing the average age in a log scale was done to emphasize the contribution from the younger stellar populations, taking into account that the variations of the SSP spectra with age do not scale linearly with age. Average ages calculated in a linear scale will result in different values, as well as average ages weighted by mass.

The range of metallicity templates was constrained by removing the very metal-poor model, nevertheless the outskirts of the nuclear disc still presents a low metallicity, and at the same time has a large contribution of recent formed stellar populations (Fig.~\ref{fig:stellar_average_light}).
Possible solutions to this issue was addressed by \citet{e.emsellem2021}, showing that the inclusion of younger SSP models result in an only marginally less metal-poor stellar population.

\subsection{Star formation history recovery}

Using the stellar population model fitted to the observed spectra with \texttt{pPXF} at the bin level, we model the SFH and derive the surface density star formation rate.
To do that, we apply the mass-to-light ratio in the $V$-band for the MILES library, the spectrum luminosity, and the light-weights for each stellar population model:
\begin{gather}
    m_{j}(t, [\rm{M/H}], [\alpha/\rm{Fe}]) = \sum \limits_{i} \left( \mathrm{\dfrac{M_{\star+remn}}{L_V}} \right)_{i}  w_{i, \rm{norm}}^{\rm{LW}} \mathrm{L}_{j, \mathrm{V},\odot},
\end{gather}
where for each bin $j$, SSP model $i$ in the (t, [M/H], [$\alpha$/Fe]) grid, $w_{i, \rm{norm}}^{\rm{LW}}$ is the normalised light-weight, corresponding to the light fraction, $\left( \mathrm{M_{\star+remn} / L_V} \right)_{i}$ is the mass-to-light ratio on the $V$-band, and $\mathrm{L}_{j, \mathrm{V},\odot}$ is the observed spectrum luminosity in solar units.

\begin{figure*}
\includegraphics[width=\textwidth]{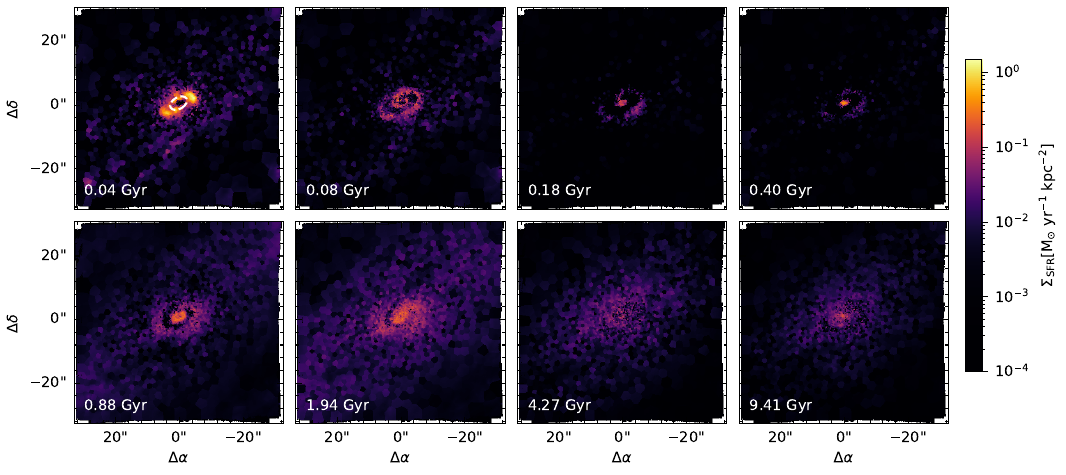}
\caption{Star formation rate surface density of the recovered SFH ($\Sigma_{\rm{SFR}}$) in $M_{\odot}$~yr$^{-1}$~kpc$^{-2}$. The nuclear disc contour (white dashed ellipse in the upper left panel) is shown with the dimensions from \citet{dimitria.gadotti2020} in the upper left panel.}
\label{fig:sfr_panels}
\end{figure*}

This results in SFH in mass fraction, allowing us to compute the mass increment in solar masses for each interval of age of the model by integrating $m_{j}(t, [\rm{M/H}], [\alpha/\rm{Fe}])$ along the metallicity and $[\alpha/\rm{Fe}]$ and consequently derive the SFR as a function of time.
In Fig.~\ref{fig:sfr_panels}, the star formation rate surface density of the recovered SFH is shown as a function of the lookback time.

This figure is consistent with the inside-out scenario of nuclear disc formation. 
In an early stage of the galaxy evolution, the star formation is active first in the centre and then progressively moves outwards, as described in \citet{bittner2020}.
The most recent episodes of star formation are located in the outskirts of the nuclear disc, whose contour is shown overlaid in the same figure, using the parameters from \citet{dimitria.gadotti2020}.
The star formation activity within the disc is detected in early stages of the galaxy evolution, being connected to the bar formation and allowing the bar age dating through suitable sophisticated methods \cite[see][]{desa-freitas2023a, desa-freitas2023,sa-freitas2025}.

It is important to note that Fig.~\ref{fig:sfr_panels} show the current location of stars with different ages, but this does not necessarily coincide with the location where they were formed.
Where the star formation episode took place is subject to galaxy dynamics and evolution, and it is not necessarily where it is seen now.
In addition, early episodes of star formation may be washed out by the model resolution for older populations.
On the other hand, the derived stellar population parameters in regions with an important contribution of stars younger than 30~Myr, the lower limit of the  MILES SSP, might be biased, and may require specific stellar population models tailored to these systems \citep[e.g.][]{leitherer1999}.

\section{Gas content}
\label{sect:gas_content}

\begin{figure*}
\centering
\includegraphics[width=\textwidth]{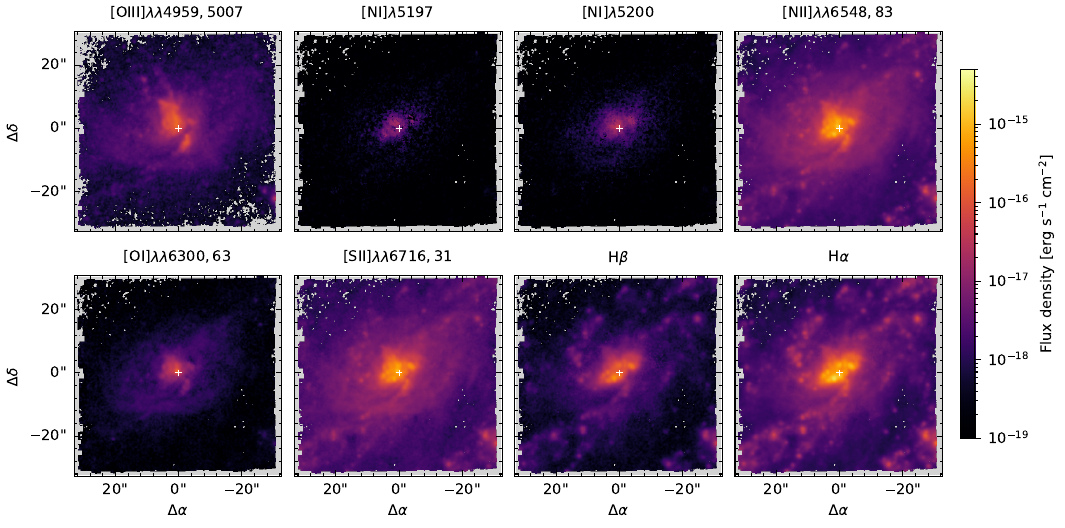}
\caption{Modelled flux of the emission lines, including all lines listed in Table~\ref{tab:emission_lines} but \ion{He}{I}, which could not be detected in the FoV. The fluxes are the combination of the individual Gaussian model of each component. The FoV centre is marked by the central crosses.}
\label{fig:emission line_panel}
\end{figure*}

The overall emission line flux along the FoV is presented in Fig.~\ref{fig:emission line_panel}, where the combined flux is shown, obtained from the multiple components with AIC selection.
The structure observed from [\ion{N}{II}], [\ion{S}{II}], and Balmer lines are quite coherent, and its most notable intensity comes from the region where the nuclear disc is located.
On the other hand, the [\ion{O}{III}] emission shows up the outflow-like emission, with a biconical shape, where the jet arises from the centre of the nuclear disc, while its counterpart is observed immediately below it.

\subsection{Extinction}

The dust extinction was estimated both from the stellar population and gas.
The stellar dust extinction was obtained as described in Sect.~\ref{sect:stellar_dust_fitting}, while the gas extinction was estimated using the Balmer decrement from H$\alpha$ and H$\beta$ which is relatively insensitive to density and temperature variations compared with its sensitivity to ISM dust.
Assuming the case B recombination, with a typical temperature $T=10^4$~K and electron density $n_e=10^2$~cm$^{-3}$, results in an intrinsic H$\alpha$/H$\beta$ ratio of 2.863 \citep{osterbrock2006}.
The colour excess $E(B-V)$ is computed adopting $\kappa(\lambda)$ as the \citet{calzetti2000} extinction curve, resulting in the following relation \citep{dominguez2013, momcheva2013}:
\begin{equation}
\centering
    E(B-V) = \dfrac{-2.5}{\kappa(\lambda_{\rm{H}\alpha}) - \kappa(\lambda_{\rm{H}\beta})} \log_{10} \left[\dfrac{(\rm{H}\alpha / \rm{H}\beta)_{obs}}{2.86}\right],
\end{equation}
where the observed fluxes of the Balmer lines are the combination of the multiple components modelling.

\begin{figure}
\centering
\includegraphics[width=\columnwidth]{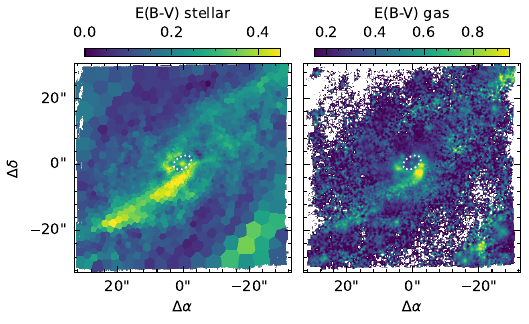}
\caption{Colour excess $E(B-V)$ from stellar component and gas. We show the colour excess that was estimated with \texttt{pPXF} (left), and the gas $E(B-V)$ was obtained from the Balmer decrement (right). The nuclear disc ellipse dimension \citep[white dotted][]{dimitria.gadotti2020} is marked in the plot.}
\label{fig:extinction_maps}
\end{figure}

The regions with enhanced extinction obtained from the stellar and nebular components trace roughly the same structures as seen in Fig.~\ref{fig:extinction_maps}, which are in agreement with the position of the bar dust lanes, and the nuclear disc, with part of the spiral arm showing in the bottom right. 
In the figure, regions where the amplitude over noise of both H$\alpha$ and H$\beta$ do not achieve a minimum of $3\sigma$ are masked, and the same applies to spaxels where the estimated extinction was close to zero and fluctuated to non-physical values.

However, the estimated colour excesses might be biased by some factors.
Despite adopting a homogeneous intrinsic ratio in the whole FoV for estimating extinction, the intrinsic $\rm{H}\alpha / \rm{H}\beta$ ratio can be affected by collisional excitation and present deviations from the adopted value.
In the extended partially ionised regions produced by a hard ionising field, collisional excitation can contribute to the Balmer line ionisation.
The contribution is higher to H$\alpha$ than for the other Balmer lines, and as a consequence, in regions with suitable conditions (e.g. AGN narrow-line regions), the intrinsic Balmer ratio is $\rm{H}\alpha / \rm{H}\beta\approx3.1$ \citep{osterbrock2006}.
NGC~613 presents a combination of ionisation mechanisms \citep{davies2017}, and the homogenous value for the intrinsic ratio adopted is just for the sake of simplicity.
On the other hand, the absolute value of the stellar population $E(B-V)$ can be biased, as estimated using overlapping pointings by \citet{e.emsellem2021}.

\subsection{Ionisation mechanism}
\label{sect:ionisation}
In NGC~613, the gas is ionised by a combination of  radiation fields, including young stars and a low-luminosity AGN.
Different sources of radiation can lead to distinct physical condition in the ionised regions, which are reflected in the emission lines.
Diagnostic diagrams are then devised to characterise the dominant ionising source of the ISM, based on ratios of emission lines.

\begin{figure}
\centering
\includegraphics[width=\columnwidth]{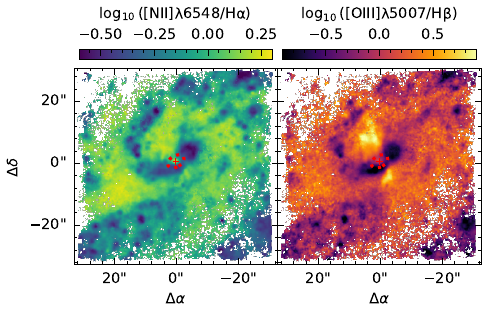}
\includegraphics[width=\columnwidth]{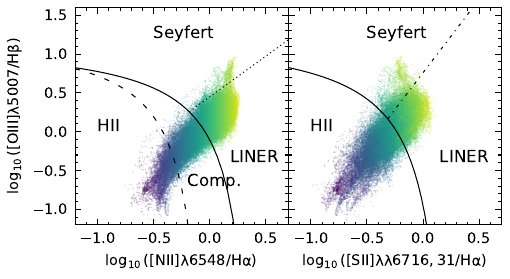}
\caption{Ionisation mechanism classification with emission line ratios. 
We present spatially resolved [\ion{N}{II}]/H$\alpha$ and [\ion{O}{III}]/H$\beta$ ratios, as a proxy for the ionisation sequence (top left and top right, respectively). In the bottom row, the BPT diagrams are shown, both colour-coded according to the top left panel scale. 
In the BPT diagrams, the dashed line corresponds to the \citet{kauffmann2003} demarcation, the continuous to \citet{kewley2001}, the dotted line to the results from \citet{kevinschawinski2007}, and dash-dotted from \citet{kewley2006}. 
The nuclear disc dimension is marked by the central ellipse into the top panels.}
\label{fig:bpt}
\end{figure}

One of the most widespread diagrams with optical lines is the BPT \citep{baldwin1981}, employing a few strong emission line ratios which include [\ion{O}{iii}]$\lambda5007$, [\ion{N}{ii}]$\lambda6548$, [\ion{S}{ii}]$\lambda\lambda6716,31$ and the Balmer lines.
In the top panel of Fig.~\ref{fig:bpt}, we show the [\ion{N}{ii}]/H$\alpha$ ratio in the FoV employing the sum of the emission line components.
This emission line ratio is commonly used in diagnostic diagrams and trace the ionisation sequence from \ion{H}{ii} to AGN/LINER.
In the two bottom panels, the BPT diagram is shown with two combinations of emission line pairs, the left- and right-hand panels exhibit [\ion{N}{ii}]/H$\alpha$ and [\ion{S}{ii}]/H$\alpha$ vs [\ion{O}{iii}]/H$\beta$, respectively, both colour coded with the same scale as the top left panel. 

The BPT diagrams reveal two main branches for large samples of galaxies: the star formation sequence and the AGN sequence. 
In the star-forming sequence, the [\ion{N}{ii}]/H$\alpha$ increases and the [\ion{O}{iii}]/H$\beta$ decreases with a more metal-rich ISM gas-phase, on the other hand, for the AGN branch, both [\ion{N}{ii}]/H$\alpha$ and [\ion{O}{iii}]/H$\beta$ increase towards larger contribution of AGN/LINER in the gas ionisation \citep{kauffmann2003}.
The transition between the two sequences takes place in the diagram locus of more metal-rich and massive galaxies, with a mixing contamination of AGN contribution.
Several empirical and theoretical curves have been proposed in BPT which are intended to define the nature of the ionisation source, based on its location on the diagram.
In the bottom panels, the continuous curves are defined by \citet{kewley2001}, being the theoretical limit of emission line ratios in starburst, and the dashed line in the left-hand panel is the empirical ratio of star-forming galaxies from \citet{kauffmann2003}.
Finally, the region between these two curves encompasses objects classified as with composite spectra. 
Beside the star-forming/AGN demarcations, we include the boundary between Seyfert and LINERs in the left-hand panel, which follows the definition from \citet{kevinschawinski2007} and in the right-hand panel the limit established by \citet{kewley2001}.

The diagnostic diagrams for NGC~613 display a mix of ionisation mechanisms. 
The top panels of Fig.~\ref{fig:bpt} exhibit structures with varying ionisation mechanisms, with a smooth transition in between, spanning from the bottom of the star-forming sequence to the top of the AGN branch in the Seyfert and LINER regions of the BPT.
Disentangling the ionisation mechanisms for NGC~613 was a case study of a novel method in \citet{davies2017}, where they identify a mixing of ionising sources.
The nuclear disc is dominated by star-forming regions, where young stars may be the responsible for the ionisation, providing UV photons. 
At the same time, the location at the bottom of the sequence indicates the high metallicity in the gas-phase, which is also indicated by the ISM oxygen abundance measurements (see Sect.~\ref{sect:gas_metallicity}).
Furthermore, the outflow is observed above, and, to a lesser extent, it is seen also below the nuclear disc, and may be related to the Seyfert activity, presenting high [\ion{N}{ii}]/H$\alpha$ and [\ion{O}{iii}]/H$\beta$ ratios.
In the neighbouring of the outflow, the [\ion{O}{iii}]/H$\beta$ decreases but a high [\ion{N}{ii}]/H$\alpha$ is still present, thus placing this region in the LINER locus of the BPT.
In the following, we further explore the activity classification with the emission components in isolation.

Next, we employ the WHAN diagram \citep{cidfernandes2010, cidfernandes2011} to identify the main ionisation source with regard to each emission line component.
This diagram is devised intending to allow the inclusion of the sources with milder emission, which often do not fulfil the requirements of signal intensity to be reliably measured.
At the same time, it is meant to be a more economical approach in terms of signal requirement, by replacing weaker lines such as [\ion{O}{iii}] and H$\beta$ by H$\alpha$.

The use of the WHAN diagram here is quite convenient due to the line intensities and the number of components detected.
Firstly, the lines employed on WHAN are often the brightest ones, thus providing robust measurements.
This is important because the secondary emission line components tend to be weaker with respect to the first component, despite the requirements of detection level applied, followed by model selection with AIC.
Secondly, we were able to detect a different number of components for low and high ionisation lines at spaxel level (see Fig.~\ref{fig:aic_selection}). 
Thus, the use of the BPT for each component, in isolation, would be difficult because of the unmatching number of components for the [\ion{O}{iii}]/H$\beta$ ratio.
Furthermore, we stress that the lines from high and low ionisation group employed may have differences in their kinematics, which may represent an additional limitation to use of this diagram.

\begin{figure*}
\centering
\includegraphics[width=\textwidth]{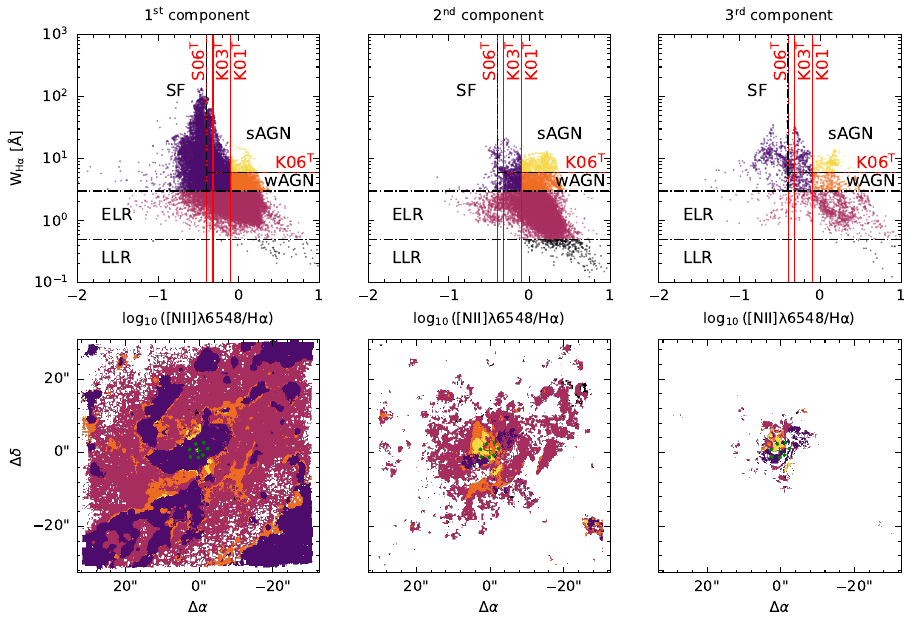}
\caption{WHAN diagram for each emission line component.
The left, middle, and right columns employ the results from the first, second and third components, respectively.
The WHAN diagrams are shown in the top row. The black dash-dotted lines delimit the regions according to the dominant ionisation mechanism: star forming (SF), strong and weak AGN (sAGN and wAGN, respectively), emission line-less retired and line-less retired \citep[ELR and ELR, respectively,][for discussion]{cidfernandes2011, herpich2016}.
The continuous red lines represent transpositions of the BPT done by \citet{cidfernandes2010} from \citet[K01$^{\rm T}$]{kewley2001}, \citet[K03$^{\rm T}$]{kauffmann2003}, \citet[S06$^{\rm T}$]{stasinska2006}, and \citet[K06$^{\rm T}$]{kewley2006}.
The vertical red lines demarcate the boundary of emission explained by young stellar ionisation (left of the straight lines), in contrast with emission that requires other sources of ionisation, and the horizontal red line separates AGN and LINER emission.
In the bottom row, each FoV panel is coloured according to the respective upper panel.
In the bottom row panels, the nuclear disc contour, depicted as a green dotted ellipse, is displayed with dimensions taken from \citet{dimitria.gadotti2020}.}
\label{fig:whan}
\end{figure*}

The WHAN diagrams for NGC~613 are exhibited on the top row of Fig.~\ref{fig:whan}, showing respectively the first, second and third components. 
In comparison to the BPT, the [\ion{O}{iii}]/H$\beta$ ratio on the vertical axes were replaced by the equivalent widths of the H$\alpha$ ($W_{\rm{H\alpha}}$), while the [\ion{N}{ii}]/H$\alpha$ ratio is still employed on the x-axes.
With that, the diagram was modified for assessment of the ionising source power in relation to the stellar component.
In the same fashion as other diagrams, boundary lines are drawn to delimitate the emission line properties that characterise ionising sources of a given nature.
The red continuous lines are the transposed curves from BPT, provided by \citet{cidfernandes2010}.

Each vertical line establishes a limit between lines ratios resulting from ionisation by young stellar populations, being respectively from: \citet[][K01]{kewley2001}, \citet[][K03]{kauffmann2003}, and \citet[][S06]{stasinska2006}. 
The regions at the left-hand side of each line are the site for spectra majorly ionised by stellar processes, while at the right-hand side are spectra dominated by AGN or LINER emission.
The horizontal red line provides a distinction between AGN and LINER emission, with the transposed curve from \citet[][K06]{kewley2006}.

The black lines are the classifications from \citet{cidfernandes2011} and \citet{herpich2016}.
They divided the diagram into several regions, defining objects into the classes of strong AGN (sAGN), weak AGN (wAGN), star-forming (SF), ``emission line-less retired'' (ELR, $0.5<W_{\rm{H\alpha}}<3$~\AA) and ``line-less retired'' (LLR, $W_{\rm{H\alpha}}<0.5$ ~\AA), where ``retired'' classes include much of the objects classified as LINERs in the BPT, and are proposed to be explained by hot low-mass evolved
stars \citep[][for discussion]{cidfernandes2011, herpich2016}.
The bottom row of the figure shows the corresponding position of the spaxels in FoV, colour-coded according to the regions of the first row, and respectively for each component.

The analysis with the components in isolation reveal differences between the underlying ionisation source for each component.
The first component mostly presents regions dominated by the star formation in the central nuclear disc, and arcs extended from the centre to the outskirts.
The remaining are predominantly LINER according to transposed BPT curves or ``retired'' emission, with weak AGN classification most probably being an artefact.
The second component shows an arc-like structure potentially associated with the nuclear disc/ring. 
Altogether, the outflow pattern ionised by the AGN is present, and it is remarkable how its geometry resembles the main outflow proxy [\ion{O}{iii}] (Fig.~\ref{fig:emission line_panel}), notwithstanding this latter line is not included in the measurement.
In the centre vicinity, an extended LINER/retired region shows up. 
The excitation processes may be the result of the contribution of passively evolving hot old stars along with the shocks induced by the propagating outflow.
Finally, the third component also exhibits a combination of ionisation processes, potentially resulting from the mix of underlying phenomena seen in projection.

\subsection{Gas metallicity}
\label{sect:gas_metallicity}

The gas metallicity is an important indicator in the study of galaxy evolution, related to the stellar processes of ISM enrichment, gas removal and mixing in a galaxy.
In this section, we derive the gas metallicity measured through the oxygen abundance [O/H]. 
This measurement is based on calibrated relations, relating the strong emission line ratios produced as a function of the oxygen abundance.
This method is a practical estimation but also subject to degeneracies between other physical quantities such as temperature, pressure, and ionisation parameter \citep{maiolino2019}. 
The ionisation parameter is a substantial factor in deriving the gas-phase metallicity for NGC~613, given the combination of different ionisation sources in this galaxy.

\begin{figure*}
\includegraphics[width=\textwidth]{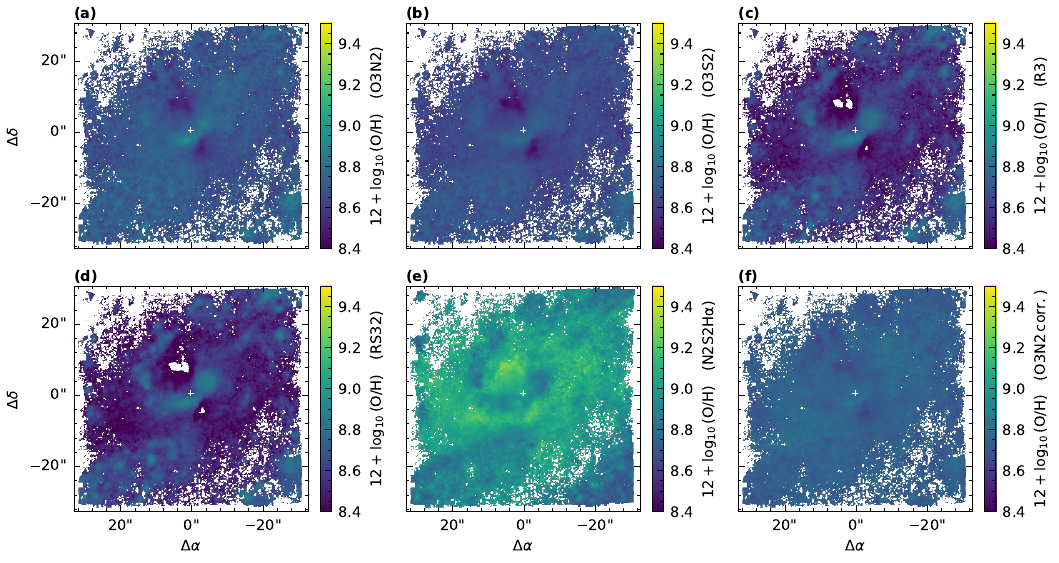}
\caption{Oxygen abundance in the gas-phase based on calibrated indexes. In the panels (a) and (b), respectively O3N2 \citep{alloin1979, pettini2004} and O3S2 \citep{curti2020} are exhibited; (c) shows the gas metallicity from R3, and (d) from RS32, both indexes with calibrations from \citet{curti2020}; (e) employs the index N2S2H$\alpha$ \citep{dopita2016, kumari2019}; and (f) shows a calibration from O3N2 additionally corrected for highly ionised regions \citep{kumari2019}.}
\label{fig:gas_metallicity}
\end{figure*}

Given the wavelength range encompassed by the observation, several strong lines commonly adopted to metallicity estimations are available.
However, the absence of strong near ultraviolet lines, like $[\ion{O}{II}]\lambda3727$ and $[\ion{Ne}{III}]\lambda3870$, prevents the use of some indexes.
In Fig.~\ref{fig:gas_metallicity}, we present $12 + \log_{10}(\rm{O/H})$ obtained from several indexes, using the integrated profiles of emission lines, resulting from the combination of the multiple components.
The first panel (a) shows the oxygen abundance from $\rm{O3N2} \equiv \log_{10}([\ion{O}{III}]\lambda5007/H\beta)-\log_{10}([\ion{N}{II}]\lambda6584/H\alpha)$ \citep{alloin1979, pettini2004}, and for $12 + \log_{10}(\rm{O/H}) = 8.73 - 0.32 \times \rm{O3N2}$ , valid for $-1 < \rm{O3N2} < 1.9$ \citep{pettini2004}.
The index O3N2 is robust against attenuation due to the pair of lines being close in the spectral axis, but it is sensitive to the ionisation parameter and N/O ratio \citep{maiolino2019}.
In this panel, the apparent metal-poor outflow immediately above and below the nuclear disc in the FoV centre is remarkable.

The surprising low metallicity of the outflow indicated with the previous index motivated the use of other indexes.
The index O3S2 is shown on panel (b), with the polynomial regression from \citet{curti2020}, and adopting the definition $\rm{O3S2} \equiv \log_{10}([\ion{O}{III}]\lambda5007/H\beta)-\log_{10}([\ion{S}{II}]\lambda\lambda6716,31/H\alpha)$.
This index is similar to O3N2, given the close ionisation potentials of S$^+$ and N$^+$, and the gas-phase metallicity obtained presents the same metal-poor outflow.
The panels (c) and (d) show the oxygen abundances computed from $\rm{R3} \equiv \log_{10}([\ion{O}{III}]\lambda5007/H\beta)$ and $\rm{RS32} \equiv \log_{10}([\ion{O}{III}]\lambda5007/H\beta)+\log_{10}([\ion{S}{II}]\lambda\lambda6716,31/H\alpha)$, respectively.
Both indexes are related to oxygen abundances according to the calibration from \citet{curti2020}. Neither of them vary monotonically with metallicity, and both have two branches at lower abundances ($12 + \log_{10}(\rm{O/H})\lessapprox8$).
The abundances obtained by these last two indexes are similar to each other, but show larger differences between the disc region and the outskirts than the abundance maps obtained from O3N2 and O3S2. 
In addition, the line ratios in the region dominated by the outflow are outside the regressions, where they could not provide an estimation of the metallicity.
In panel (e), we show the results employing the N2S2H$\alpha \equiv \log_{10}([\ion{N}{II}]\lambda6584/[\ion{S}{II}]\lambda\lambda6716,31)+0.264\log_{10}([\ion{N}{II}]\lambda6584/\rm{H}\alpha$ \citep{dopita2016}; this calibration intends to reduce the ionisation parameter dependence with $[\ion{N}{II}]/\rm{H}\alpha$.
We estimate the oxygen abundance from N2S2H$\alpha$ with the Equation~3 from \citet{kumari2019}.

Finally, in panel (f), we employ the re-calibration proposed by \citet{kumari2019}, intended to be suitable to regions with a higher ionisation degree, like diffuse ionised gas (DIG) and low ionisation regions (LIER) that otherwise would be more susceptible to measurement bias of oxygen abundance.
The O3N2 calibration, usually derived for \ion{H}{II} regions, is corrected taking into account the [\ion{O}{III}]/H$\alpha$ ratio, to reduce the bias associated to the ionisation parameter.
The use of this corrected calibration produces a higher estimate of metallicity in the outflow region, and a lower estimate for the nuclear disc in comparison to the calibration used in panel (a).
Nevertheless, to a lesser extent, the outflow is still more metal-poor than the disc.
It cannot be ruled out that, despite the correction that minimises the bias, there may be potential limitations associated with the use of indexes in the metallicity estimation that could result in the  low metallicity of the gas in the outflow being an artifact.

\subsection{Gas kinematics}
\label{sect:gas_kinematics}

\begin{figure*}
\centering
\includegraphics[width=\textwidth]{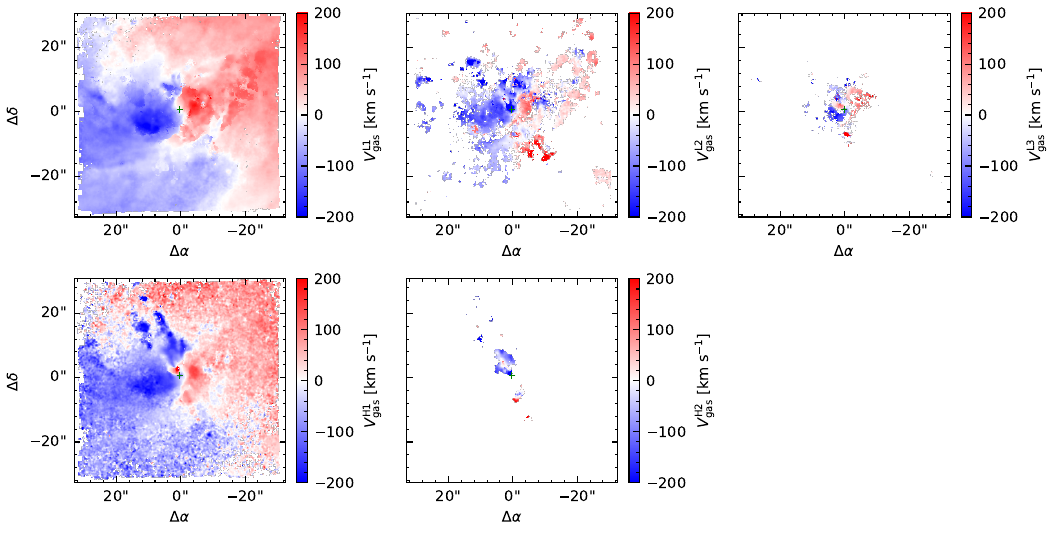}
\caption{Gas kinematics for each component and emission line group. Top row: The LOSV of the gas, for the emission line group of low ionisation potential (Table \ref{tab:emission_lines}), respectively for the first ($V^{\rm{L1}}_{\rm{gas}}$), second ($V^{\rm{L2}}_{\rm{gas}}$), and third ($V^{\rm{L3}}_{\rm{gas}}$) components. Bottom row: The LOSV of the gas as in the top row, but for the first ($V^{\rm{H1}}_{\rm{gas}})$ and second ($V^{\rm{H2}}_{\rm{gas}}$) component of the high ionisation potential group.}
\label{fig:velocity-panel}
\end{figure*}

The kinematics was modelled divided into two groups of emission lines, with low- and high-ionisation  potential (L and H upper indexes respectively, see also Table~\ref{tab:emission_lines}).
In Fig.~\ref{fig:velocity-panel}, we present the LOSV of the gas emission lines ($V_{\rm{gas}}$), with the corresponding velocity dispersions shown on Appendix~\ref{sect:dispersion}.
The use of distinct groups is proven to be substantial to reveal the particularities of each group that otherwise would be washed out in favour of the kinematics of a single physical phenomenon.

When fitting with multiple components, deviations from a pure Gaussian in the emission line profile that otherwise would appear as asymmetries can be modelled and disclose several underlying physical phenomena.
The interpretation of all the features fitted is not straightforward and might be challenging.
The first association that can be made is that the narrowest component in the emission is associated to the regular dynamics of the  gas in the disc, with the $V^{\rm{L1}}_{\rm{gas}}$ being similar to the rotation pattern exhibited by the stellar disc.
While for $V^{\rm{H1}}_{\rm{gas}}$ and $V^{\rm{H2}}_{\rm{gas}}$ the kinematics of the outflow is traced, the high-ionisation lines seem to follow the same pattern as $V^{\rm{L1}}_{\rm{gas}}$ in the outskirts.
The $V^{\rm{L2}}_{\rm{gas}}$ also exhibits features that may be associated with outflow, and at the same time, this component presents higher velocity dispersions (see Fig.~\ref{fig:dispersion}) than its corresponding first component ($V^{\rm{L1}}_{\rm{gas}}$).
Finally, the $V^{\rm{L3}}_{\rm{gas}}$ has a less straightforward interpretation, and this component may not correspond to a single phenomenon.

One of the interesting aspects in the study of the gas kinematics is to understand how it can be affected by stellar structures such as the bar.
In the stellar disc, the gas dynamics is dictated by the same gravitational potential that constrains the stellar orbits, however, it is a complex potential given the presence of several stellar structures.
These stellar structures result in orbits that are overlaid or self crossing, in the non-inertial, rotating frame of the bar.
Gas parcels following these orbits may lead to shocks in the gas that removes angular momentum and induces inflows.
These shocks are specially expected due to the bar, given the families of orbits that support its induced orbits. The shocks occur within the bar dust lanes, leading to the compression of the gas in the leading edge of the bar.

Differences of the kinematics of the baryonic components may be observable due to the distinct nature of stars and gas interactions: the non-collisional stellar motion in contrast with the gas hydrodynamical behaviour.
These differences can be highlighted when the gas motion is observed in relation to the stellar component.
In particular, we intend to put in evidence the gas kinematics deviations from the circular motion of the disc.

We then model a simple rotating disc, from the stellar kinematics, with \texttt{Kinemetry} \citep{krajnovic2006}.
The steps adopted follow the procedure outlined in \citet{kolcu2023}, which we refer for an in-depth description of the process.
As a result, $V_{\rm{circ}}$ will be the best-fit model of the line-of-sight circular velocity of the NGC~613 disc.

The disc of NGC~613 is rotating anticlockwise, assuming the dust lanes are leading the bar \citep{sormani2023}, with trailing spiral arms (Fig.~\ref{fig:ngc613}), and the inclination of the stellar disc is~$39^{\circ}$ in relation to the plane of the sky \citep{buta2015}.
By combining these propositions, it can be inferred that the disc parcel located in the bottom region of the FoV is closer to the observer, whereas the disc parcel situated in the top region is farther away.

\begin{figure*}
\includegraphics[width=\textwidth]{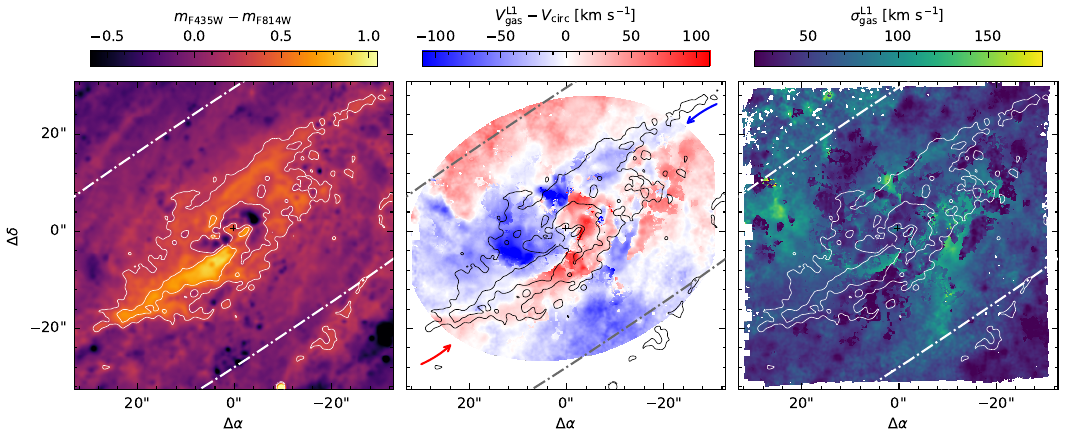}
\caption{Bar dust lane contour, gas velocity in relation to stellar, and gas velocity dispersion. In the left-hand panel, the HST colour index from the filters F435W and F814W is shown, with the regions more affected by extinction used as a proxy for the location of the bar dust lanes. The gas velocity of the first component is shown in relation to the modelled circular velocity in the middle panel, exhibiting the non-circular motions of the gas, with portions of the gas along the bar dust lane, receding (red arrow) and approaching (blue arrow) in the observer's frame of reference, and moving towards the centre of the galaxy. The right-hand panel shows the velocity dispersion of the gas. In the middle and right-hand panels, the contour of the first panel is shown, in addition to the bar contour (also shown in the left-hand panel) with the generalised ellipse parameters from \citet{kim2014}.}
\label{fig:gas_kin_dust}
\end{figure*}

Given that the bar dust lanes correspond to the offset bar regions with higher gas density and enhanced starlight absorption \citep{athanassoula1992}, we build a colour map from HST images to create contours that highlight the bar dust lanes and nearby regions, as shown on the first panel of Fig.~\ref{fig:gas_kin_dust}.
In addition, in all the panels of the figure, the bar contour (dash-dotted lines) is shown as a generalised ellipse \citep{athanassoula1992}, with a projected bar semi-major axis of 80~arcsec, $\rm{PA}=34^{\circ}$, ellipticity $\epsilon = 0.7$, and the boxiness parameter $c=2.8$ \citep[taken from][]{kim2014}.

In the second panel, we show the LOSV of the gas ($V^{\rm{L1}}_{\rm{gas}}$) in relation to $V_{\rm{circ}}$, namely the residual motion, which reveals non-gravitational effects induced in the gas kinematics, resulting in non-circular motions.
With the residual motion, we observe an evidence of gas inflow along the bar dust lanes.
The dust lanes are located along the diagonal of the FoV. The closer one (bottom of the FoV) seems to have a higher concentration of dust in relation to the furthermost (top).
That is suggested both by $E(B-V)$ (Fig. \ref{fig:extinction_maps}), and colour index (Fig.~\ref{fig:gas_kin_dust}).
Evidence of inflow is observed along both dust lanes. 
In the bottom one, the non-circular motion can be observed from the bottom left corner of Fig.~\ref{fig:gas_kin_dust} towards the galaxy centre (red). The same is observed for the other dust lane (blue).

Part of the gas coming through the bottom dust lane seems to overshoot the centre and move towards the upper right corner, and this overshooting gas may produce sprays that  get shocked with the other dust lane \citep{sormani2019, kim2024}.
Again, this behaviour is shown on the kinematics of the gas coming from the top dust lane to get shocked with the bottom one, but presenting a remarkable gradient in the velocity field.
Gradients in the line-of-sight velocity are observed perpendicularly to the dust lane. In particular, in the bottom dust lane, a strong gradient is observed, matching a region with enhanced velocity dispersion in the third panel of the Fig.~\ref{fig:gas_kin_dust}.
Finally, in the upper left and bottom right regions of the second panel in this figure, parallel to the bar ellipse and outermost to the dust lane's contours, the residual motion of the gas presents a clockwise rotation, indicating that the gas is trailing the stellar disc in this region in relation to the line-of-sight.

\begin{figure}
\centering
\includegraphics[width=\columnwidth]{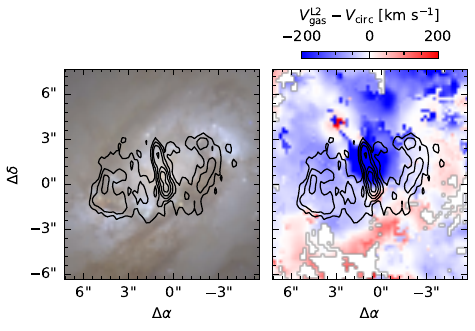}
\caption{Gas kinematics in the secondary component and radio jet in the nucleus of NGC~613. Left panel: HST image with VLA radio contour. Right panel: gas kinematics in the secondary component, overlaid with the VLA contour.}
\label{fig:gas_vla}
\end{figure}

The gas emission in the second component of the low-ionisation group traces the kinematics of energetic phenomena.
This component is characterised by broader emission lines compared to the first component (Fig.~\ref{fig:dispersion}), and are often shifted in relation to the stellar LOSV.
In Fig.~\ref{fig:gas_vla}, we emphasise the low-ionisation second component LOSV, overlaid with radio data contour \citep{hummel1992, falcon-barroso2014} from VLA C-band (4.86~GHz).
The VLA data exhibits a linear radio jet composed of three blobs with $\rm{PA}=12^\circ$.
In the same region, our emission line modelling points to emission blueshifted with a peak in the $V^{\rm{L2}}_{\rm{gas}}$ up to $\sim500$~km~s$^{-1}$ in relation to stellar disc velocity.

\subsection{Outflow}
\label{sect:outflow}

One of the most remarkable features observed from nebular emission in NGC~613 is the [\ion{O}{III}] morphology (Fig.~\ref{fig:emission line_panel}).
This emission line was modelled in the same group of \ion{He}{I}, although only [\ion{O}{III}] was detected.
According to the AIC selection (Fig.~\ref{fig:aic_selection}), up to two components are required to model the emission line profile, whose LOSV are $V^{\rm{H1}}_{\rm{gas}}$ and $V^{\rm{H2}}_{\rm{gas}}$.
The [\ion{O}{III}] flux indicates the occurrence of an ionised outflow with a biconical pattern, composed by a blueshifted and redshifted side.
The blueshifted outflow is seen in projection above the nuclear disc, while the counter redshifted outflow appear just below it.
The blueshifted side is visible in front of the galactic stellar disc, whereas the redsifhted one is visible behind it, and its observed flux is more attenuated by dust.

From the spatially resolved [\ion{O}{III}], we can evaluate some outflow properties, as for example the outflow mass.
To estimate the mass, according to the procedure described in \citet{carniani2015}, it is required to employ the $[\ion{O}{iii}]\lambda5007$ luminosity, the electron density $n_e$ and the oxygen abundance relative to solar, such that the mass of the ionised outflow is given by:
\begin{multline}
\label{eq:outflow_mass}
    M_{\rm ion} = 0.8\times10^8 M_{\odot}\\
    \times \left(\dfrac{C}{10^{[O/H]}}\right)  \left(\dfrac{L[\ion{O}{iii}]}{10^{44}~\rm{erg}~s^{-1}}\right)
    \left(\dfrac{n_e}{500~\rm{cm}^{-3}}\right)^{-1}.
\end{multline}
where $C$ is the ``condensation factor'', i.e. the ratio between the volume-averaged electron density squared and the volume-averaged squared electron density $C=\langle n_e \rangle^2 / \langle n_e^2 \rangle$, we assume $C=1$ under the physical condition assumptions described in \citet{cano-diaz2012}.
Furthermore, we employ the oxygen abundance in solar units $\rm{[O/H]}$, estimated from the O3N2 index (Sect.~\ref{sect:gas_metallicity}) and with the solar oxygen abundance $\log \epsilon_O\approx 8.86$ \citep{centeno2008}.

The electron density was obtained from the $[\ion{S}{II}]\lambda\lambda6716,31$ ratio ($R_{[\ion{S}{II}]}$), adopting the relation from \citet{sanders2016}.
This measurement saturates for the doublet ratio $R_{[\ion{S}{II}]}<0.7$ and $R_{[\ion{S}{II}]} >1.45$, with $n_e$ varying between $\sim50$ and $2\,000$~cm$^{-3}$ within these limits \citep{kakkad2018}.
Near the limits, small variations in the ratio lead to large changes in the estimated $n_e$, and in these conditions the measurement is particularly sensitive to background noise.
In the outskirts of the FoV, with moderate S/N, we have observed more noisy measurements, which we interpret as artefacts.
We then adopt an average of the electron density, which was computed in the central kiloparsec, resulting in \mbox{$n_e\approx112$~cm$^{-3}$}.
The properties of the outflow are then derived, following the procedures outlined by \citet{smethurst2019, smethurst2021}. 
With Equation~\ref{eq:outflow_mass}, we estimated the outflow mass as \mbox{$\log_{10} M_{\rm ion} = 4.97\pm0.01~M_{\odot}$}, with the value of the uncertainties representing only a lower limit, result of the standard propagation, since it neither include the spatial covariance, which is challeging for MUSE \citep[][]{bacon2017}, nor systematics errors.

The bulk velocity of the outflow \citep[$v_{[\ion{O}{iii}]}$;][]{smethurst2021}, is measured using the maximum value from $V^{\rm H1}_{\rm gas}$ and $V^{\rm H2}_{\rm gas}$ in relation to the stellar velocity component, where we employ the 0.1$^{\rm{st}}$ and 99.9$^{\rm{th}}$ quartiles of the $V_{\rm{gas}} - V_{\star}$ distribution to be robust against outlier values.
In addition, the spaxels with the maximum values of gas velocities are in good agreement with the farthermost regions of the outflow cone in relation to the galaxy centre ($\Delta \delta\approx 25\arcsec$ and $\Delta\alpha\approx12$).
Then, we adopt these points to characterise the outflow extent, resulting in $r_{\rm max}\approx2.8\pm0.2$~kpc. 
Therefore, the timescale of the outflow is estimated from the bulk outflow velocity and from the ionised outflow scale \citep{smethurst2021}:
\begin{equation}
\label{eq:timescale_outflow}
    t_{\rm out} = \left(\dfrac{r_{\rm max}}{\rm km} \right)\left(\dfrac{v_{[\ion{O}{iii}]}}{\rm{km~yr^-1}} \right)^{-1} [{\rm yr}],
\end{equation}
resulting in an estimate of $t_{\rm out} \approx 2.1\pm0.3$~Myr.

Next, the mass outflow rate was estimated with \citep{smethurst2021}:
\begin{equation}
\label{eq:mass_outflow_rate}
    \dfrac{\dot{M}_{\rm ion}}{M_{\odot}~\rm{yr}^{-1}} = B \left(\dfrac{M_{\rm ion}}{M_{\odot}} \right)\left(\dfrac{t_{\rm{out}}}{\rm{yr}} \right)^{-1},
\end{equation}
with $B=1$ \citep[see][for discussion]{smethurst2021}, resulting in an average $\dot{M}_{\rm ion} \approx0.04\pm0.01~M_{\odot}~\rm{yr}^{-1}$.
Finally, we compute the kinetic energy outflow rate $\dot{E}_{\rm out} = 0.5 \dot{M}_{\rm ion} v_{[\ion{O}{iii}]}^2$, which allows to investigate the underlying physical mechanism driving the outflow \citep{smethurst2021}, with our computations estimating 
$\dot{E}_{\rm out} = 2.4 \times 10^{40}$~erg~s$^{-1}$.

\section{Discussion}
\label{sect:discussion}

\subsection{Spectral fitting techniques}
The process of full-spectrum fitting has been developed through the last few decades, increasing the quality and quantity of properties derived.
With the algorithms devised for this task, it is possible to derive the stellar population kinematics, dust extinction and recover the SFH by making use of several approaches.
There is a fast and active development of new packages dedicated to this task, and updates in the already available and widespread tools.
Despite the huge success of studies employing these tools, the further developments indicate there is room for improvements and implementation of novel routines.
This is specially true for datacubes, where the number of spectra to be fitted and the physical connection between spaxels add new levels of complexity to the problem.

Regarding the emission line fitting with multiple components, much effort has been done to tackle this problem, which present a solution space with local minima, and are subject to good choices of initial conditions or to computational expensive global optimisation.
In particular for the large modern IFU surveys, with the increasing order of magnitude of the number of spectra processed, this becomes particularly challenging.
Further, the varying environmental conditions due to mixed physical processes across the galaxy prevent a suitable choice of initial guesses for the whole FoV.
The codes and methods to deal with these problems are also in fast development.
With the multiple components, many concurrent ionisation phenomena can be detected, but identifying spatially coherent phenomena is also challenging, remarkably in cases where the organisation of the components according to velocity dispersion may not be physically meaningful, with the employment of codes for clustering being required \citep[e.g. acorns;][]{henshaw2019}.

In this work, we implement a pipeline for the fitting process, combining steps already used in different studies with novelty approaches.
We employ the optimised mask, which are specially relevant for cleaning bad pixel, skylines, or artefacts, all of them at spaxel/bin level, yielding cleaner spectra to a more robust modelling.
Moreover, the field normalisation has proven to be a simple but effective practice in setting the regularisation parameter, then allowing a single choice of parametrisation to the whole FoV, which ultimately will lead to well-regularised SFH recovery across the FoV.

Regarding the emission lines, we employ a process that aims to fit multiple Gaussian components to the observed line profiles.
For that, global optimisation, robust against local minima, is used in a binned solution.
Next, a spaxel level solution is obtained, taking advantage from the binned one, and employing an outlier detection algorithm along with RBF interpolation to produce initial guesses.
Finally, the model selection with AIC is adopted to prevent overfitting models.
All together, these processes lead to the identification of multiple components,  a result of several concurrent physical processes, where inflow and outflow, and AGN or star-formation driven ionisation can be studied.

\subsection{The effects of bars on the gaseous content in galaxies and AGN activity}

The non-circular motions we observe in the gas are evidence of inflow induced by the stellar bar, which is well known both from theoretical studies and observations \citep[][]{athanassoula1992, piner1995, delorenzo-caceres2013, shin2017, sormani2023, kolcu2023}.
This inflow is related to the hypothesis of the bar-AGN relation, where a large scale galactic bar could favour the transport of gas to the central kiloparsec and then help to feed the supermassive black hole.
This scenario has received support from observational studies \citep{knapen2000, laine2002, p.coelho2011, oh2012, melaniegalloway2015, santiagoalonso2018, silva-lima2022, garland2024}.
In addition, models suggested that the gas can get stalled in the ILR, in a scale of a few hundred parsecs \citep{piner1995, shin2017, mattiac.sormani2018}.
In this context, inner scale structures may contribute to momentum removal and inflow to inner scales of the galaxy, for example through inner bars \citep[][]{delorenzo-caceres2013} and nuclear spiral structures \citep[][]{a.audibert2019}.
However, other structures like the boxy/peanuts can play a role in the gas inflow by reducing its rate \citep{fragkoudi2015, fragkoudi2016}.

Nevertheless, other dynamic perturbations, pair-pair interaction, can play a role in angular momentum removal from the gas in the ISM, consequently inducing gas transport and triggering AGN activity \citep{saral.ellison2019} or nuclear star-formation \cite{scudder2012}.
With studies suggesting that the combinations of companion interaction in bar potential may amplify the AGN activity \citep{alonso2024}.
Also, bars seem to enhance star-formation and nuclear activity in denser environments \citep{sanchez-garcia2023}.
In addition, other mechanisms which are not dynamically related are suggested to provide fuel to the AGN, for example stellar related ones \citep{riffel2024}.
Nevertheless, the combination of concurrent AGN feeding mechanisms, besides the distinct timescales for gas to be transported by the bars and episodes of AGN activity, jointly might make it difficult to find a potential bar-AGN relation.

\subsection{The nuclear outflow in NGC~613}
In NGC~613, in addition to the non-circular motions along the dust lanes, we observe another phase of the gas cycle related to the ionised outflow.
The outflow is traced mainly by the [\ion{O}{III}] emission in two kinematical components (Fig.~\ref{fig:velocity-panel}). 
Features in the kinematics of lower potential lines seems to be also associate with the outflow, for example, on larger scales ($V^{\rm L1}_{\rm gas}$ blueshifted and above the galaxy centre in Fig.~\ref{fig:velocity-panel}) and in inner scales (aligned with the jet in Fig.~\ref{fig:gas_vla}).

Whether the mechanism that induces this outflow is caused by star formation or by AGN is a subject that has been discussed in the literature \citep{davies2017, a.audibert2019}.
With our analysis with multiple components, we are able to model the complex line profiles exhibited in the galaxy centre. 
We find that the ionisation is simultaneously the result from the combination of AGN and star formation when employing the WHAN diagnostic diagram.

Through the emission line fitting, we can determine the kinematics and characterise the ionised outflow, including estimates of the mass and the mass outflow rate. 
The $M_{\rm ion}$, $r_{\rm max}$, and $\dot{M}_{\rm ion}$ measurements span a range of values in samples of galaxies at low-redshifts in literature, allowing the comparison with the ionised outflow properties in NGC~613.
The $M_{\rm ion}$ ranges from $\approx10^{3-8}$~$M_{\sun}$ \citep{rakshit2018}.
At the same time, the outflow radius span values from 0.5 to 5.1~kpc \citep{rakshit2018}.
\citet{smethurst2021}, for example, find $\dot{M}_{\rm ion}$ between 0.12 and 0.19~$M_{\sun}~\rm{yr}^{-1}$, while \citet{smethurst2019} include outflow rates from 0.007 up to 2.99~$M_{\sun}~\rm{yr}^{-1}$, and \citet{rakshit2018} indicates $\dot{M}_{\rm ion}$ between 0.01 and 126~$M_{\sun}~\rm{yr}^{-1}$ in a census of Seyfert~1.
The outflow properties we find for NGC 613 in this study are within the reported ranges. 

The bolometric luminosity of the AGN can be estimated from \mbox{X-ray} (2-10~keV) as $L_{\rm bol,AGN}=1.6\times10^{42}$~erg~s$^{-1}$ \citep{davies2017}, which along with $\dot{E}_{\rm out}$ and $\dot{M}_{\rm ion}$ place NGC~613 in the sequence from \citet[][left and right panels from their Fig. 14]{venturi2018}, extrapolated from the relation for ionised outflows in \citet{fiore2017}.
Furthermore, $\dot{E}_{\rm out}$ leads to a coupling factor of 0.015 or $1.5\%L_{\rm bol, AGN}$.
This coupling factor is within the range of modelled values \citep[$0.5 - 10\%$,][]{dimatteo2005,hopkins2010}, indicating that the AGN is capable to power the observed ionised outflow, and in turn the outflow can possibly quench star-formation.
The most intense star formation in NGC~613 centre is occurring in the nuclear disk outskirts as suggested in the modelled SFH (Fig.~\ref{fig:sfr_panels}), and the last episodes of star formation in the nucleus of NGC~613 was estimated to take place $\sim10$~Myr ago, while the timescale of the outflow indicates a $t_{\rm out}\sim2$~Myr.
Our estimation of the outflow timescale might be limited, since the extent is comparable to the FoV size, but given the $r_{\rm max}$ in local galaxies \citep{rakshit2018}, it may not be larger than a factor of 2, which may constraint the outflow timescale.
In addition, the bulk velocity estimation try to take into account limitations of the measurement, and may result in differences regarding the adopted corrections \citep[e.g.][]{rakshit2018, smethurst2021}, which is also reflected in the other estimated properties of the outflow. 

\section {Summary and Conclusions}
\label{sect:conclusion}

In this work, we explore the gas kinematics in NGC~613, a nearby late-type barred galaxy rich in stellar substructures, which, at the same time, harbours complex activity distributed in its central regions, both from AGN and star-formation. 
To this purpose, we use MUSE IFU observations from the TIMER survey. 
We model the emission lines with multiple components, taking the underlying stellar continuum into account, developing our own dedicated analysis pipeline for this purpose. 
Our main findings are as follows:
\begin{enumerate}
    \item We highlight the effect of different normalisation approaches of the observations in the parametrisation of the regularisation, with the scalar field normalisation providing substantial improvements in the SFH recovery.
    \item We employ an approach to fit multiple Gaussian components in the modelling of emission lines, tested to be robust against local minima, which combines global optimisation and Levenberg–Marquardt algorithms available on \texttt{pPXF} with model selection provided by AIC.
    \item The gas-phase in the central region of NGC~613 presents a mix of ionisation mechanisms. We derive multiple components in the emission lines, separated at the spaxel level, meaning that we can analyse each component and its underlying ionising physical process in isolation.
    \item The integral field spectroscopy allow distinguishing between several spatially distributed physical process, with the distinct conditions of the ISM imposing challenges in some measurements. As an example, we find a surprising low metallicity in the outflow region in relation to the disc. This disparity is partially reduced when employing calibrations from literature where additional corrections for the ionisation degree are adopted.
    \item With the multiple components, along with the partition of emission lines of distinct ionisation potential in different groups, we are able to capture the kinematics of the gas undergoing different physical processes. With that, we could observe evidence of inflow along the bar dust lanes, and the kinematics of the gas in the outflow.
    \item The remarkable ionised gas outflow present in the centre of NGC~613 is traced mainly through the [\ion{O}{iii}] flux; with this emission line, the mass outflow rate was estimated to be 0.04~$M_{\odot}$~yr$^{-1}$.
\end{enumerate}

The stellar component in a galaxy may interact with the gas-phase, among other means, through its gravitational potential, and as a consequence affecting processes such as AGN and star formation.
The gravitational effect may be even more drastic due to strongly non-axisymmetric components, for example, stellar bars.
With the several nuclear stellar structures seen in galaxies, the direct relation between a stellar bar and star-formation or AGN may be elusive, due to the intricate interaction of multiple stellar components and differences in the timescales of the physical phenomena.
In a future study, we will employ the methodology developed here to study a sample of galaxies observed with MUSE, composed of barred and unbarred galaxies. 
This will help us in understanding the role of bars in the gas-phase dynamics and in the process of building reservoirs of gas in sub-kiloparsec scales.

\section*{Acknowledgements}

The authors are grateful to the anonymous reviewer for their constructive suggestions, which significantly strengthened this manuscript.
This study was financed in part by the Coordenação de Aperfeiçoamento de Pessoal de Nível Superior - Brasil (CAPES) - Finance Code 88887.637633/2021-0.
This research was supported by the Munich Institute for Astro-, Particle and BioPhysics (MIAPbP), which is funded by the Deutsche Forschungsgemeinschaft (DFG, German Research Foundation) under Germany´s Excellence Strategy – EXC-2094 – 390783311.
LASL was supported by Conselho Nacional de Desenvolvimento Científico e Tecnológico (CNPq) via project 200469/2022-3/SWE. Based on observations collected at the European Southern Observatory under ESO programme 097.B-0640(A).
LASL thanks Anelise Audibert for VLA reduced data.
DAG is supported by STFC grants ST/T000244/1 and ST/X001075/1. 
This work used the DiRAC@Durham facility managed by the Institute for Computational Cosmology on behalf of the STFC DiRAC HPC Facility (www.dirac.ac.uk). The equipment was funded by BEIS capital funding via STFC capital grants ST/K00042X/1, ST/P002293/1, ST/R002371/1 and ST/S002502/1, Durham University and STFC operations grant ST/R000832/1. DiRAC is part of the National e-Infrastructure.
L.P.M. thanks FAPESP (grant 2022/03703-1) and CNPQ (grant 307115/2021-6) for partial funding of this research.
T.Kolcu also acknowledges the financial support from The Leverhulme Trust.
PC acknowledges support from Conselho Nacional de Desenvolvimento Cient\'ifico e Tecnol\'ogico (CNPq) under grant 310555/2021-3
FF is supported by a UKRI Future Leaders Fellowship (grant no. MR/X033740/1). This work was supported by STFC with grants ST/T000244/1 and ST/X001075/1.
T.Kim acknowledges support from the Basic Science Research Program through the National Research Foundation of Korea (NRF) funded by the Ministry of Education (No. 2022R1A4A3031306, No. RS-2023-00240212).
J.F-B acknowledges support from the PID2022-140869NB-I00 grant from the Spanish Ministry of Science and Innovation.
AdLC acknowledges financial support from the Spanish Ministry of Science and Innovation (MICINN) through RYC2022-035838-I and PID2021-128131NB-I00 (CoBEARD project).
JMA acknowledges the support of the Viera y Clavijo Senior program funded by ACIISI and ULL and the support of the Agencia Estatal de Investigación del Ministerio de Ciencia e Innovación (MCIN/AEI/10.13039/501100011033) under grant nos. PID2021-128131NB-I00 and CNS2022-135482 and the European Regional Development Fund (ERDF) ‘A way of making Europe’ and the ‘NextGenerationEU/PRTR’.

\section*{Data Availability}

Raw and reduced MUSE data used in this work can be accessed via the ESO Science Archive Facility.



\bibliographystyle{mnras}
\bibliography{biblio} 




\appendix

\section{Velocity dispersion maps}

\begin{figure*}
\includegraphics[width=\textwidth]{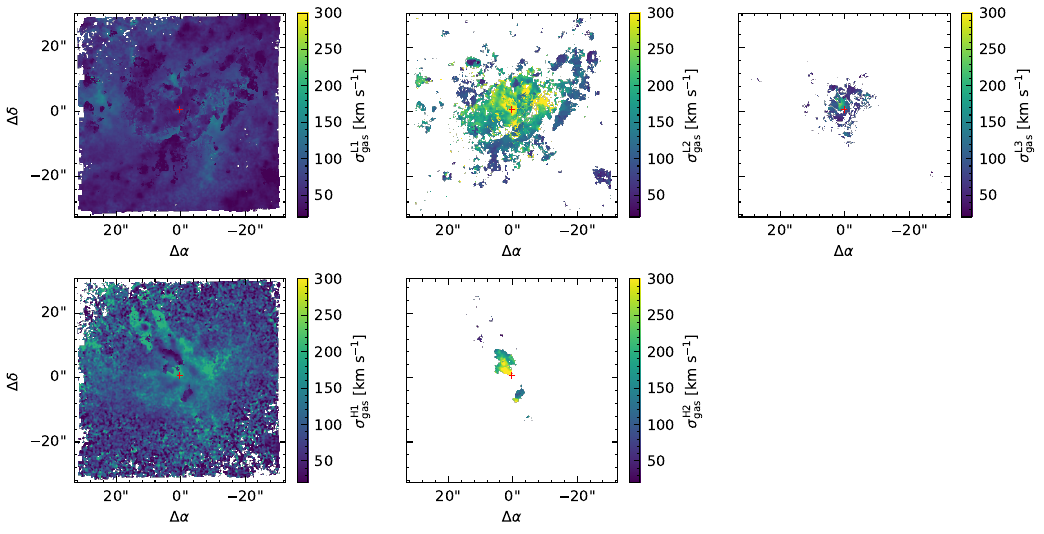}
\caption{Velocity dispersion maps. The velocity dispersion corresponding to each of the panels in Fig.~\ref{fig:velocity-panel}.}
\label{fig:dispersion}
\end{figure*}

The velocity dispersion of each component is shown on Fig.~\ref{fig:dispersion}.
Some jumps in the transition from regions with different component are some artefacts from the fit. 
In these transitions, with the change in the number of components, the width of the Gaussians slightly increase or decrease, as the number of components decrease or increase respectively. 
\label{sect:dispersion}


\bsp	
\label{lastpage}
\end{document}